# An Atlas of Star-Forming Galaxy Equivalent Widths

Short title: Star-Forming Galaxy Equivalent Widths Atlas

Helen Meskhidze[1*] and Chris T. Richardson[1*]

1. Physics Department, Elon University, Elon, NC, 27244, USA

**Abstract**

We present an atlas of starburst galaxy emission lines spanning 10 orders of magnitude in ionizing flux and 7 orders of magnitude in hydrogen number density. Coupling SEDs from Starburst99 with photoionization calculations from Cloudy, we track 96 emission lines from 977 Å to 205 μm which are common to nebular regions, have been observed in H II regions, and serve as useful diagnostic lines. Each simulation grid displays equivalent widths and contains ~$1.5 \times 10^4$ photoionization models calculated by supplying a spectral energy distribution, chemical abundances, dust content, and gas metallicity (ranging from 0.2 $Z_\odot$ to 5.0 $Z_\odot$). Our simulations will prove useful in starburst emission line data analysis, especially regarding local starburst galaxies that show high ionization emission lines. One sample application of our atlas predicts that C IV λ1549 will serve as a useful diagnostic emission line of vigorous star formation for coming *James Webb Space Telescope* observations predicting a peak equivalent width of approximately 316 Å.

## 1. Introduction

Starburst galaxies, also known as H II galaxies or star-forming galaxies, feature strong emission lines primarily due to newly formed massive stars. The emission line diagnostic diagram displaying [O III] λ5007/ Hβ vs. [N II] λ6584/ Hα, the BPT diagram (Baldwin et al. 1981; hereafter BPT), has been remarkably successful in empirically dividing extreme star-forming galaxies from their extreme AGN counterparts. Later work added theoretical upper (Kewley et al. 2001) and lower limits for identifying star-forming galaxies along with a line on the BPT diagram dividing AGN from LINERS (Kauffman et al. 2003).

As mentioned above, along the extreme "wings" of the BPT diagram, AGN and star-forming galaxies are easily distinguished from one another. However, while starlight is often the dominant source of excitation in star-forming galaxies, several other excitation mechanisms can contribute to the production of emission lines. The most intensely star forming galaxies along the far left wing of the BPT diagram are often interacting or merging (Robaina et al. 2009). Strong shocks further contribute to the excitation of the gas in galaxies along the left wing.

Similarly, at lower ionization, classifying galaxies as star-forming or AGN becomes difficult. For these galaxies, excitation and ionization of gaseous clouds could likely be the result of starlight, non-thermal sources, or a combination of the two. Historically, the presence of a radiation field hard enough to generate photons higher than 50 eV signified excitation from an AGN. However, modern models of stellar radiation fields that incorporate Wolf-Rayet (WR) stars produce a significant number of EUV photons capable of ionizing heavy elements through many excitation states.

---

[*] emeskhidze@elon.edu, crichardson17@elon.edu, 100 Campus Drive, Elon NC 27244 (336-278-6281)

However, the presence of high ionization emission lines signifying AGN activity stands in conflict with typical classification schemes. NGC 3621 provides an example of an optically classified star-forming galaxy at low redshift that emits [Ne V] in the infrared (Satyapal et al. 2007). In the local ($z \sim 0$) neighborhood, several star-forming galaxies show weak nebular [O IV] emission with AGN activity ruled out by spatial resolved spectroscopy around the size of a starburst region (Lutz et al. 1998). Similarly, Sharzi and Brinchman (2012) found a significant number of optically classified star-forming galaxies with strong He II λ4686 emission within the Sloan Digital Sky Survey (SDSS) at $z \sim 0-0.4$.

Observations indicate a larger influence of vigorous star formation on emission line production at early times in the universe (Madau & Dickinson 2014. This leads to models which incorporate the starburst phase prior to any AGN activity (Hopkins et al. 2006). Indeed, local star-forming galaxies that exhibit characteristics of Lyman break galaxies (LBGs) can diverge from the common sequence of starburst galaxies along the BPT diagram (Stanway et al. 2014). For example, at higher redshifts of $z \sim 1-3$, emission line galaxies with high [O III] / Hβ ratios are frequently located in regions of the BPT diagram that are typically unoccupied by nearby galaxies, limiting our understanding regarding the level of contribution from star formation (Liu et al. 2008; Steidel et al. 2014). Local analogs for high-$z$ galaxies provide crucial case studies for understanding these galaxies that have higher electron densities and ionization parameters than those in the local universe (Brinchmann et al. 2008; Bian et al. 2016).

Modeling star-forming galaxies with spectral synthesis codes provides the key link to understanding the gas conditions and excitation mechanisms that are necessary to reproduce high ionization emission in starburst galaxies. A common technique for such modeling involves coupling a spectral energy distribution predicted from a population synthesis code with a photoionization code that predicts the observed spectrum. This technique has been used in a large number of previous studies (e.g. Abel et al. 2008; Levesque et al. 2010; Stark et al. 2014; Richardson et al. 2013).

The overlapping goal of many of these studies is to understand the physical parameters responsible for the variation in the emission line spectrum of the objects within a given sample. One of the most frequently used methods for fitting star-forming galaxy spectra involves assuming a single spectral energy distribution (SED) and electron density while varying the gas metallicity, $Z$, and ionization parameter, $U$, defined as:

$$U = \frac{\phi_H}{c\, n_H} \quad (1)$$

with $\phi_H$ representing hydrogen ionization photon flux [cm$^{-2}$ s$^{-1}$], and $n_H$ representing the hydrogen number density (Kewley et al. 2001). Other studies (Mas-Hesse & Kunth 1999, Charlot & Longhetti 2001, Levesque et al. 2010) have included a sensitivity studies to document the effects of an aging starburst on the traditional emission line ratios assuming $n_e = 100$ cm$^{-3}$, $\log(U) = -2.2$. The prescription of varying the cloud $U$ and $Z$ for a single age starburst has proven useful in fitting galaxy spectra with small He II / Hβ values observed in the local universe (Shirazi & Brinchmann 2012) but fails to fit to largest star forming galaxy He II / Hβ ratios (Guseva et al. 2000).

A relatively low $U$ value consistent with H II regions in our own galaxy provides a reasonably accurate model of spectra for most nearby galaxies; however, to construct a model that represents an even greater fraction of local galaxies, one must also include the higher $U$ values necessary for the [O III] / Hβ ratios found in local analogs (Richardson et al. 2013). Incorporating higher ionization parameters ($\log(U) > 0.0$) in plasma simulations has been successful in explaining infrared [Ne V], [Ne III] and [O IV] emission (Abel et al. 2008) along with [N IV] emission in the UV (Raiter et al. 2010).

A more recent interpretation for understanding star formation has resulted from applying a locally optimally emitting cloud (LOC) model to reproduce a large number of emission line ratios (Richardson et al. 2014). The central idea behind an LOC model comes from the fact that emission seen from a distant observer reflects the *cumulative* emission of all clouds around a central ionization source. As first shown by Baldwin et al. (1995), this results in powerful selection effects: we observe the emission from clouds that are optimally tuned to emit them. Since the different emission lines optimally emit under vastly different physical conditions, the locally optimally emitting cloud (LOC) model incorporates a wide range of ionizing fluxes and densities.

LOC modeling was first developed to understand the selection effects present in the broad line region (BLR) of quasars. In particular, Korista et al. (1997; hereafter K97) provided an atlas of equivalent widths ($W_\lambda$) over the LOC plane for many prominent emission lines present in quasars. This work emphasized the selection effects inherent to the BLR and set a solid foundation for subsequent modeling, including the coupling of the accretion disk to the inner torus (Goad, Korista & Ruff 2012) and the estimation of central black hole masses (Negrete et al. 2012).

In this paper, we use an LOC methodology to focus on the sensitivity of typical photoionization model parameters in producing higher ionization emission lines and notoriously weak emission lines. Our results will provide observers with an understanding of what conditions could produce anomalous emission in star-forming galaxies in the low-$z$ universe, aid in distinguishing between possible excitation mechanisms, supply baseline grids for LOC integration modeling (Richardson et al. 2016), and inform next generation surveys about the best possible emission line wavelengths to probe starburst galaxies.

We follow in the footsteps of K97 in documenting the selection effects associated with observations by providing an atlas of starburst galaxy equivalent widths. Specifically, we are guided by the following questions: 1. *What are the inherent selection effects present in unresolved starburst galaxy observations?* 2. *What physical conditions are necessary to produce strong high ionization emission lines assuming photoionization via starlight?* 3. *To what degree can star clusters contaminate emission line observations of galaxies classified as AGN?*

To probe the answers to our guiding questions, we present a massive suite of plasma simulations that are based on a LOC methodology and feature only photoionization. Our simulations span a large range of $U$, $Z$, grain abundance and star-formation history (SFH). Unlike previous work, we do not explicitly specify the ionization parameter in our calculations. Instead, our standard simulation grids include a vast range of $\phi_H$ and $n_H$, which reveals variations in the emission line properties present in clouds with similar ionization parameters. We limit our analysis to low-$z$ or typical Orion conditions, but also explore simulations involving more extreme conditions. We thus present equivalent widths for 92 emission lines covering wavelengths from EUV to the FIR. A full list of the 167 emission lines predicted in our simulations is given in Appendix A. Our choice in emission lines is guided by not only strong lines, but also weaker, and thus less commonly measured lines, along with lines that have diagnostic value (e.g. $n_e$, $T_e$, SFR, etc.).

In §2 we present our spectral energy distributions that were generated by a population synthesis code and were used as input for our plasma simulations. In §3 we present our baseline model along with the physical characteristics of clouds in the LOC plane. We follow this up with a comprehensive set of equivalent widths, covering a large range of wavelengths, and discuss many of the prominent features. We elaborate on the differences across the LOC plane associated with gas metallicity, SFH, and dust content in §4. The implications of our results on local galaxies and future *James Webb Space Telescope* observations are presented in §5; and finally, in §6, we summarize our results and outline avenues for future work.

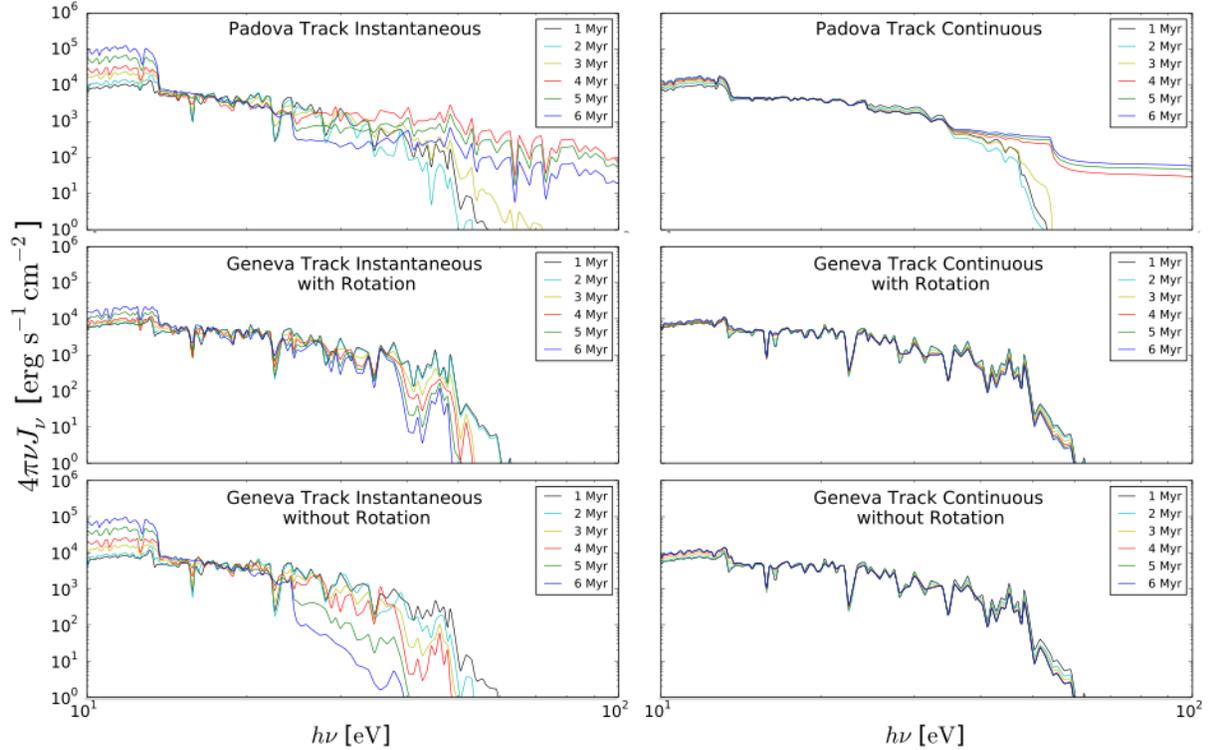

**Figure 1.** Starburst99 SEDs. The synthetic spectra generated by Starburst99 to serve as input for our photoionization simulations. The left panels show spectra from star clusters undergoing as single episode of star formation, while the right panels show spectra from star clusters continually forming new stars. Each row represents a different evolutionary track. All stars in these simulations have solar metallicity

## 2. Population Synthesis Synthetic Spectra

For generating our incident spectral energy distributions, we use the code Starburst99 (Leitherer et al. 1999). We explore the Padova track evolutionary sequence with Asymptotic Giant Branch (AGB) stars (Bressan et al. 1993) and the Geneva evolutionary sequence with zero rotation and 40% break up velocity (Leitherer et al. 2014). For each track, we include the Pauldrach/Hillier model atmospheres (Pauldrach et al. 2001; Hillier & Miller 1998). We assumed a Kroupa broken power law initial mass function (IMF; Kroupa 2001) with mass intervals of 0.1 $M_\odot$ to 0.5 $M_\odot$ and 0.5 $M_\odot$ to 100 $M_\odot$, which are the default values for a Starburst99 simulation.

Each evolutionary sequence of Starburst99 uses either a continuous or instantaneous SFH. Our instantaneous starbursts assume a fixed mass of $10^6$ $M_\odot$, while our continuous starbursts assumed a star formation rate of 1 $M_\odot$ yr$^{-1}$, both of which are the default parameters for a Starburst99 simulation.

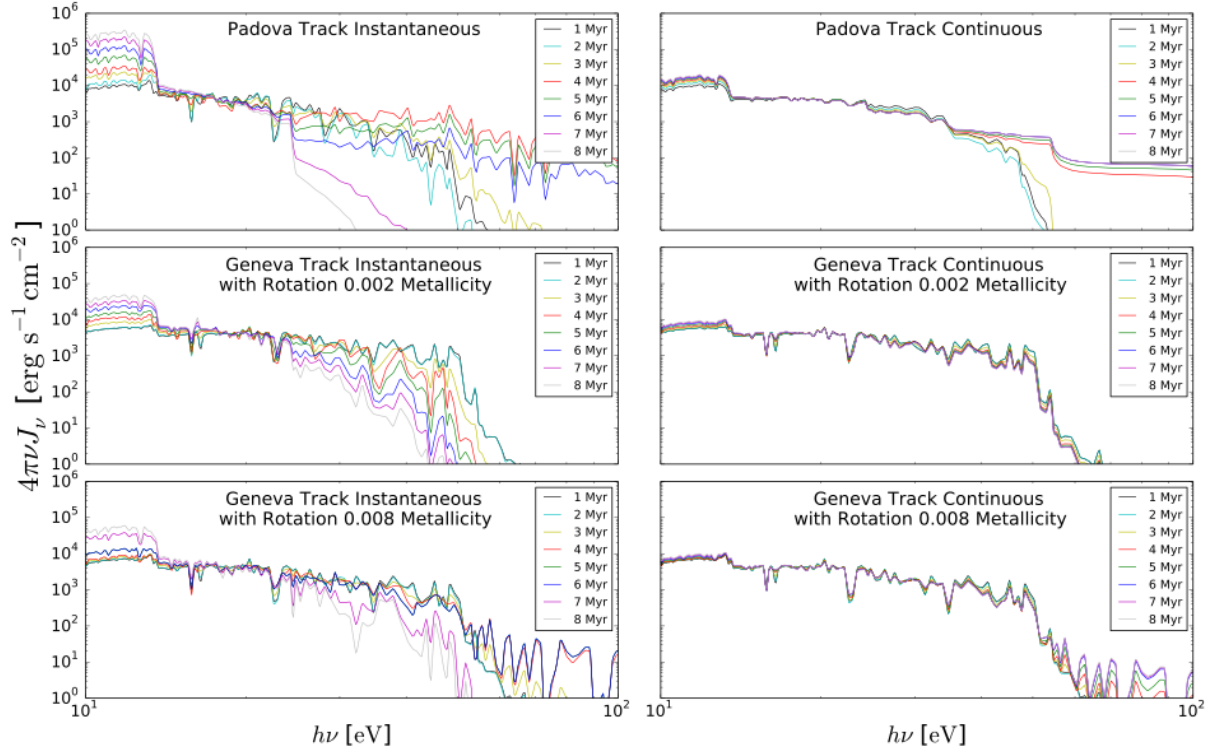

**Figure 2**. Starburst99 SEDs varying the metallicity. The bottom panel shows spectra from a 0.4 $Z_\odot$ star cluster and the middle panel displays spectra from a 0.1 $Z_\odot$ star cluster. The top panel is repeated from Figure 1 to show that despite incorporating stellar rotation with the Geneva evolutionary tracks, the hardest SED results from a star cluster with a continuous SFH reaching steady state and following the Padova evolutionary track

We investigate the sensitivity of the SED to two additional parameters: SFH (including stellar population age) and stellar metallicity. The greatest effect comes from the SFH (Figure 1), with stellar metallicity introducing less noticeable changes to the overall spectrum. In particular, the effects of stellar metallicity were less noticeable when the Geneva track continuous evolution model was adopted (Figure 2).

Figure 1 displays the spectra from star clusters with instantaneous SFHs in the left panel and the spectra from star clusters with continuous SFHs in the right panels. The bottom two rows of spectra are distinguished by differences in stellar rotation following the Geneva evolutionary track, while the top row features spectra following the Padova AGB evolutionary track.

As evident in Figure 1, stellar rotation affects the radiation field in a number of ways. Specifically, rotating stars will spend a longer amount of time on the main sequence, and will have higher effective temperatures and luminosities than non-rotating stars (Levesque et al. 2012). Rotation also increases the number of WR stars by enhancing mass loss, which allows stars of lower mass to enter a WR phase.

Despite these effects, the overall hardness of the ionizing spectrum from solar metallicity stars with continuous SFH is fairly similar for non-rotating and rotating stars as shown in Figure 1. At subsolar metallicities however, the effects of rotation become more apparent. Figure 2 displays the Padova AGB track and Geneva Rotation track spectra for both SFHs, however the Padova track star clusters have solar metallicity while the Geneva track star clusters have subsolar metallicities (0.1 $Z_\odot$ and 0.4 $Z_\odot$). At lower metallicity, the star cluster takes 10-20% longer to reach steady state, the point when the radiation field ceases to evolve (Leitherer et al. 2014). At 0.4 $Z_\odot$, the effects of rotation on the hardness of the spectrum become much more apparent. As the star cluster becomes even more metal poor, stars do not have sufficient mass loss to enter the WR phase and thus the hardness of the spectrum deceases, relative to the spectrum emitted from 0.4 $Z_\odot$ stars (Figure 2). In spite of rotation resulting in a greater number of higher energy photons, the steady state Padova AGB track SED at 5 Myr or older produces the hardest ionizing spectrum, which can be seen by comparing the FUV and EUV intensities in Figure 1 and Figure 2. We note that binary evolution also significantly increases EUV intensities (Stanway et al. 2014); however, we have only included secular, or isolated, evolution in this work for simplicity. As noted in Shirazi & Brinchmann (2012) and Jaskot & Ravindranth (2016), binary evolution causes an increase in the EUV flux for a longer period of time, however the peak of the flux does not significantly differ from the SED we have chosen for our baseline model described in the next section.

**3. Baseline Model**

Our baseline model assumes characteristic values for various parameters of starburst regions in Cloudy (Ferland et al. 2013). Below, we justify these choices as well as discuss the characteristics of the emission lines produced by adopting such parameters. We first detail the input parameters of our model, including the assumed abundances and boundary conditions. Next, we explain features of the model, including temperature contours and grain sublimation points. Lastly, we explore the variation of emission line contours across our grid (in UV, optical, and IR) for emission lines that show to positive equivalent widths in our baseline model.

3.1 Input Parameters
*3.1.1 Spectral Energy Distribution*

As discussed in the introduction, we are interested in reproducing observed high ionization potential emission lines and probing the conditions inferred by recent work (e.g. Guseva et al. 2000, Thuan & Izotov 2005, Kewley et al. 2013). We are guided by the findings of Abel & Satyapal (2008) and Shirazi & Brinchmann (2012), who investigated local starburst galaxies ($z < 0.6$) and found [Ne V] 14.3 μm and He II λ4686 emission lines respectively. The production of high-energy photons requires substantial WR star populations, which are not generated by many of the stellar population models we considered here.

In order to assess the effects of a WR population, we compare the peak $W_\lambda$ of higher ionization potential emission lines across the LOC plane. For example, we find that the peak $W_\lambda$ of [Ne V] λ3426 is about 5 times greater for the Padova continuous evolution track than the Padova instantaneous evolution track. We note that the Geneva track instantaneous evolution model at $Z = 0.008 Z_\odot$ and 5 Myr results in marginally more emission from most emission lines (e.g. [O II] λ3727 and [O III] λ5007 emission is slightly higher, peaking at around 1.25 times the baseline model peak value), but less emission from high ionization emission lines. In particular, the Padova continuous model at 5 Myr predicts generates [Ne V] λ3426 with positive equivalent width, along with [Ne V] λ24.1 μm and [Ne V] λ14.3 μm emission roughly three times greater than the emission predicted using the Geneva track instantaneous evolution model at $Z = 0.008 Z_\odot$ at 5 Myr with rotation. For this reason, and simplicity, we adopt the Padova AGB continuous evolution track SED at 5 Myr as our baseline model.

In principle, we note that the choice of the stellar metallicity for the SED should be determined on a case-by-case basis for each galaxy, since the history of the galaxy plays a significant role. In one scenario, setting the stellar and gas-phase metallicities equal to one another assumes that the stars formed from the same gas that they later excite. In another scenario, setting the stellar and gas-phase metallicities separately assumes that recent galaxy mergers mixed interstellar gas with different stellar populations, as is the case with blue centered galaxies (Stark et al. 2013). In the following sections, one should note that the results from our baseline model do not strictly adhere to one of these scenarios.

*3.1.2 Boundary Conditions*

Our simulations begin at the illuminated face of the cloud. The two main stopping conditions we adopt are total hydrogen column density, $N(H)$, and electron temperature, $T_e$. Our simulations stop when $N(H)$ exceeds $10^{23}$ cm$^{-2}$ because Cloudy becomes optically thick to Compton scattering. In a later section (§ 4.1), we explore the sensitivity of relaxing this condition. Additionally, we stop our models when the $T_e$ falls below 4000 K because gas hotter than 4000 K is required to produce collisionally excited optical and UV lines.

We note that in our high metallicity simulations, we observed a pocket of little emission in the bottom left corner of the LOC plane (low $\phi_H$, low $n_H$) for many of the emission lines we track when the two above stopping conditions were imposed. To ensure that we were reaching the ionization front of the cloud in our high-metallicity simulations, we thus removed the temperature stopping condition. Instead, we stop our high metallicity simulations when $n_e / n_H < 0.5$.

*3.1.3 Abundances*

Since dust is a ubiquitous feature of H II regions, we include it in our baseline model. Carbonaceous and silicate grains are included in the grid wherever dust sublimation does not occur, according to the dust and gas phase abundances given in Baldwin et al. (1991). Full dust abundances are based on Orion (Baldwin et al. 1991) and given by number relative to hydrogen (Table 1). In grid locations where total dust sublimation occurs, solar abundances (Grevesse et al. 2010) are adopted (§4.4). For the dust free models, we adopt solar abundances across the plane without any grains, which are given by number relative to hydrogen in Table 1.

**Table 1.** Abundance sets for Orion nebula (dusty simulations) and Solar (dust-free simulations) by number, relative to hydrogen, in the log.

| Element | Orion Nebula | Solar |
|---|---|---|
| H | 0.00 | 0.00 |
| He | -1.02 | -1.07 |
| Li | -10.27 | -11.17 |
| Be | -20.00 | -10.84 |
| B | -10.05 | -9.52 |
| C | -3.52 | -3.79 |
| N | -4.15 | -4.61 |
| O | -3.40 | -3.53 |
| F | -20.00 | -7.66 |
| Ne | -4.22 | -4.29 |
| Na | -6.52 | -5.98 |
| Mg | -5.52 | -4.62 |
| Al | -6.70 | -5.77 |
| Si | -5.40 | -4.71 |
| P | -6.80 | -6.81 |
| S | -5.00 | -5.10 |
| Cl | -7.00 | -6.72 |
| Ar | -5.52 | -5.82 |
| K | -7.96 | -7.19 |
| Ca | -7.70 | -5.88 |
| Sc | -20.00 | -9.07 |
| Ti | -9.24 | -7.27 |
| V | -10.00 | -8.29 |
| Cr | -8.00 | -6.58 |
| Mn | -7.64 | -6.79 |
| Fe | -5.52 | -4.72 |
| Co | -20.00 | -7.23 |
| Ni | -7.00 | -6.00 |
| Co | -8.82 | -8.03 |
| Zi | -7.70 | -7.66 |

*3.1.4 Hydrogen Density and Incident Ionizing Flux*

The limits we impose over the LOC plane are based on physical limits rather than observational ones. The limits for $n_H$ in our baseline grid are based on the low density limit (LDL) and the critical density, $n_{crit}$, values of the emission lines we tracked. The LDL for most emission lines is ~$10^0$ cm$^{-3}$, so we chose this as our lowest $n_H$ value. The upper $n_H$ limit of $10^6$ cm$^{-3}$ was chosen to roughly correspond to the $n_{crit}$ value for common optical emission lines used for classifying star forming galaxies (e.g. [O I] λ6300, [N II] λ6584), however we note that the overwhelming majority of H II regions will fall within the $10^2$ - $10^4$ cm$^{-3}$ range.

The incident ionizing flux, $\phi_H$, typically is not defined explicitly in photoionization simulations, but rather through a version of ionization parameter:

$$q = \frac{\phi_H}{n_H} \quad (2)$$

Leveque et al. (2010) adopt a range of $7 \leq \log(q) \leq 8.6$, which, when including their range of hydrogen density, $1 \leq \log(n_H) \leq 2$, translates to $8 \leq \log(\phi_H) \leq 10.6$. In our simulations grids, our lower limit to $\phi_H$ is set by the lower limit of Levesque et al. (2010), while the upper limit is set by the grain sublimation point, a process highly unlikely to occur and which has not been observed in local H II regions. All together, this sets our simulation grid at $8 \leq \log(\phi_H) \leq 17$, which is much broader than any other studies, but is appropriate given our focus on higher ionization emission lines.

*3.2 Physical Conditions Across the LOC Plane*

We next analyze the dependence of electron temperature on location in the $n_H$ and $\phi_H$ plane. In Figure 3, we have plotted contours of $T_e$ at the face of the gas cloud. The red lines represent increments of 0.2 dex, while the black lines represent increments of 1 dex. Note that though in Figure 3, we show temperatures that fall below the cut-off temperature of 4000 K, these have negligible contributions to the spectrum overall because they are stopped after one zone.

In Figure 3, we also show a sample of ionization parameter contours in teal. Since the ionization parameter is dependent on both $\phi_H$ and $n_H$ (as described in §3.1.4), contours of the ionization parameter create constant sloped lines on our grid. As seen in Figure 3, temperature increases with increasing $U$, reaching a peak of $10^7$ K. It should be noted that there is not a strong relationship between $n_H$ and $T_e$, but that certain heating and cooling mechanisms do change slightly along constant $U$ values (K97). This produces the slight variation in temperature along the $U$ contour lines.

Though the general trend exhibited by the temperature contours is consistent with the literature (e.g. Richardson et al. 2014, K97), we find a dip in our temperature contours around $\log(n_H) = 1$ where our models include grains. Tracking the heating at $\log(\phi_H) = 12$, we find that at $\log(n_H) = 0$, the heating is dominated by grains and He I. However at $\log(n_H) = 1$, He II and H I dominate the heating. This trend continues as $2 \leq \log(n_H) \leq 4$. The cooling at $\log(n_H) = 0$ is dominated by O VI and dust, but then at $\log(n_H) = 1$, Ne V and C IV are the dominant cooling mechanisms. Then, from $2 \leq \log(n_H) \leq 4$, O IV and O III dominate cooling again. This suggests that a change in the dominant cooling mechanism may be causing the fluctuation in temperature across the LOC plane. While interesting, this temperature fluctuation is minor and thus does not raise any concerns. In the bottom left of this figure, we superimpose the parameter space explored by three other studies: depicted in grey is the study of Levesque et al. 2010, in blue is Kewley et al. 2001, and in yellow is Moy et al. 2001.

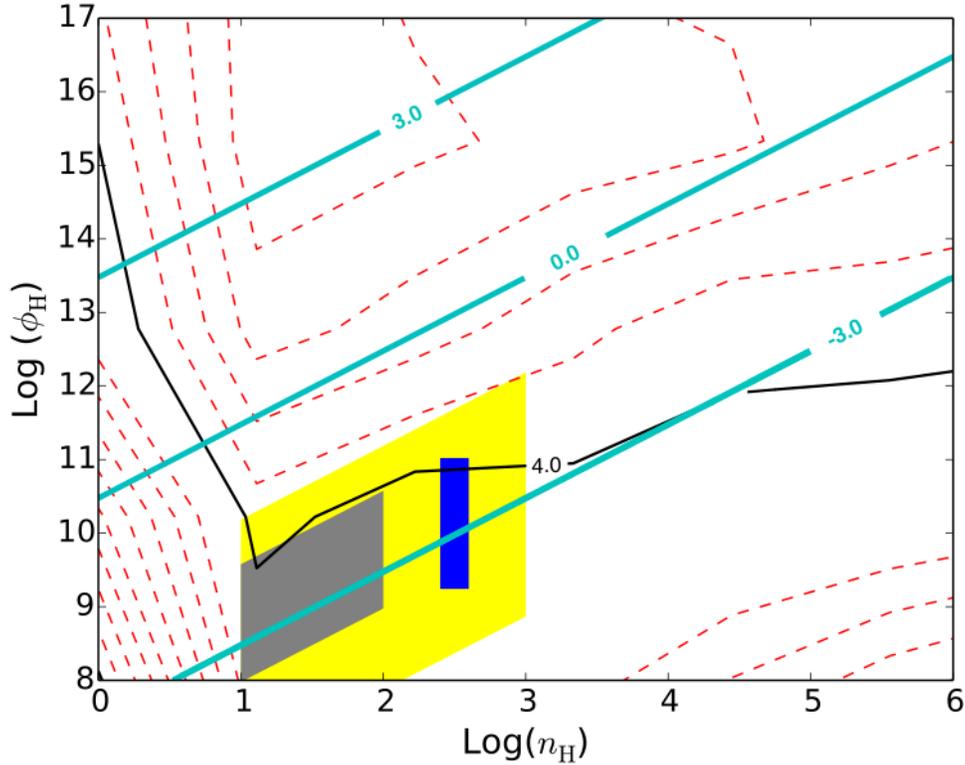

**Figure 3.** Temperature contours across the LOC plane. Contours of the electron temperature ($T_e$) and ionization parameter ($U$) at the face of the cloud as a function of hydrogen ionizing photon flux ($\phi_H$) and the hydrogen density ($n_H$). The baseline continuum was used for this model (Padova track, continuous evolution with grains and solar abundances). The teal solid lines show a few contours of the ionization parameter across the LOC plane. The red and black lines show the $T_e$ with increments of 0.2 dex and 1 dex respectively. In the bottom left of this figure, we also show the parameter space explored by the studies of Levesque et al. 2010 in grey, Kewley et al. 2001 in blue, and Moy et al. 2001 in yellow.

### 3.3 Equivalent Width Predictions

Many of the emission lines emit in a narrow range of ionization parameters. There is little emission in the bottom right corner of our grids (low $U$: high $n_H$, low $\phi_H$) where the gas is under-ionized, and even less in the top left corner (high $U$: low $n_H$, high $\phi_H$) where the gas is over-ionized. Another way to understand these general trends in emission is by analyzing optical depth at 912Å. On our grids, some of the optically thin clouds are located in the upper left corner where the temperature is $10^5 < T_e < 10^7$ K (Figure 3). Most of the emission lines we present do not emit at such high temperatures; consequently, there is little emission by most clouds in this region of our grids. Another set of optically thin clouds are located in the lower right corner of the grids, with temperatures $10^2 < T_e < 10^3$ K. Note that our simulations are truncated at temperatures below 4000 K (§3.1.2). While some optical and infrared emission lines emit in these extreme regions, the efficiency of reprocessing is generally very low making emission lines weak.

Contrarily, optically thick clouds are located diagonally across the grids from the lower left corner to the upper right. The reprocessing efficiency in optically thick clouds is much higher for most of our emission lines. This is because ionizing photons in optically thick clouds face a greater probability of absorption before escaping, which leads to emission lines emitting more strongly. Notably, along the top of the ridge (when moving from the optically thick to the optically thin region), the optical depth drops off severely, sometimes by 6 orders of magnitude. This corresponds to $1 < \log(U) < 2$. Along the bottom of the ridge ($\log(U) \approx -4$), this drop off is much less severe. Finally, though located diagonally across the LOC plane, the divisions between optically thin and thick clouds are not along constant ionization contour lines. The optically thick region broadens with higher $n_H$ and $\phi_H$ values.

*3.3.1 UV Emission Lines*

Figure 4a displays the equivalent widths across the LOC plane for selected UV emission lines. Collisionally excited lines, such as C IV λ1549 (Figure 4a, row e), generally show the most efficient reprocessing of the spectrum along constant ionization parameter lines, which span from low $n_H$ and low $\phi_H$ values to high $n_H$ and high $\phi_H$ values. In our simulations, this corresponds to a $\log(U) \approx -1.5$. When moving perpendicularly to the constant $\log(U) \approx -1.5$, there is a sharp decrease in $W_\lambda$ because $U$ becomes either too low (when moving orthogonally downwards) or too high (when moving orthogonally upwards). For column densities of $10^{23}$ cm$^{-2}$, one of the stopping criterion of our simulations, clouds with $\log(U) \gtrsim 0.5$ are optically thin to He$^+$ photons and so they reprocess little of the incident continuum. Many other UV emission lines exhibit these same trends, bands of emission along constant $U$ lines with sharp declines perpendicular to the constant $U$ lines.

The ratio of C III λ2297 to C IV λ1549, a dielectric recombination line and a collisionally excited line respectively, is a temperature indicator (Osterbrock & Ferland 2006, hereafter AGN3). When this ratio is low, the temperature in the nebula is high. Our baseline model predicts very little C III λ2297 (completely flat at 0.3 dex; Figure 4a, row g) emission and substantial C IV λ1549 (around 1.5 dex where C III λ2297 peaks; Figure 4a, row e), meaning that the ratio of these two emission lines is $< 10^{-2}$. The temperatures predicted are thus between 10000 K and 15000 K, which is consistent with Figure 3. Alternatively, the ratio of [C III] λ1907 to C III] λ1909 is a $n_e$ probe (AGN3; Figure 4a, row g). The lower the ratio between these two emission lines, the higher the $n_e$. This ratio is around 0.5 on our grids in the moderate density range in which [C III] λ1907 emits, but drops significantly with increased density (in the region where [C III] λ1907 ceases to emit but C III] λ1909 emits most strongly).

C IV λ1549 can be contrasted with lower ionization emission lines that are still collisionally excited, such as Mg II λ2798 (Figure 4a, row h). Since Mg II λ2798 is a lower ionization emission line, its peak of $\log(W_\lambda) \approx 2.0$ is higher than that of C IV λ1549, which peaks at $\log(W_\lambda) = 1.5$. Additionally, the peak $W_\lambda$ is shifted to a lower $U$ than that of C IV λ1549.

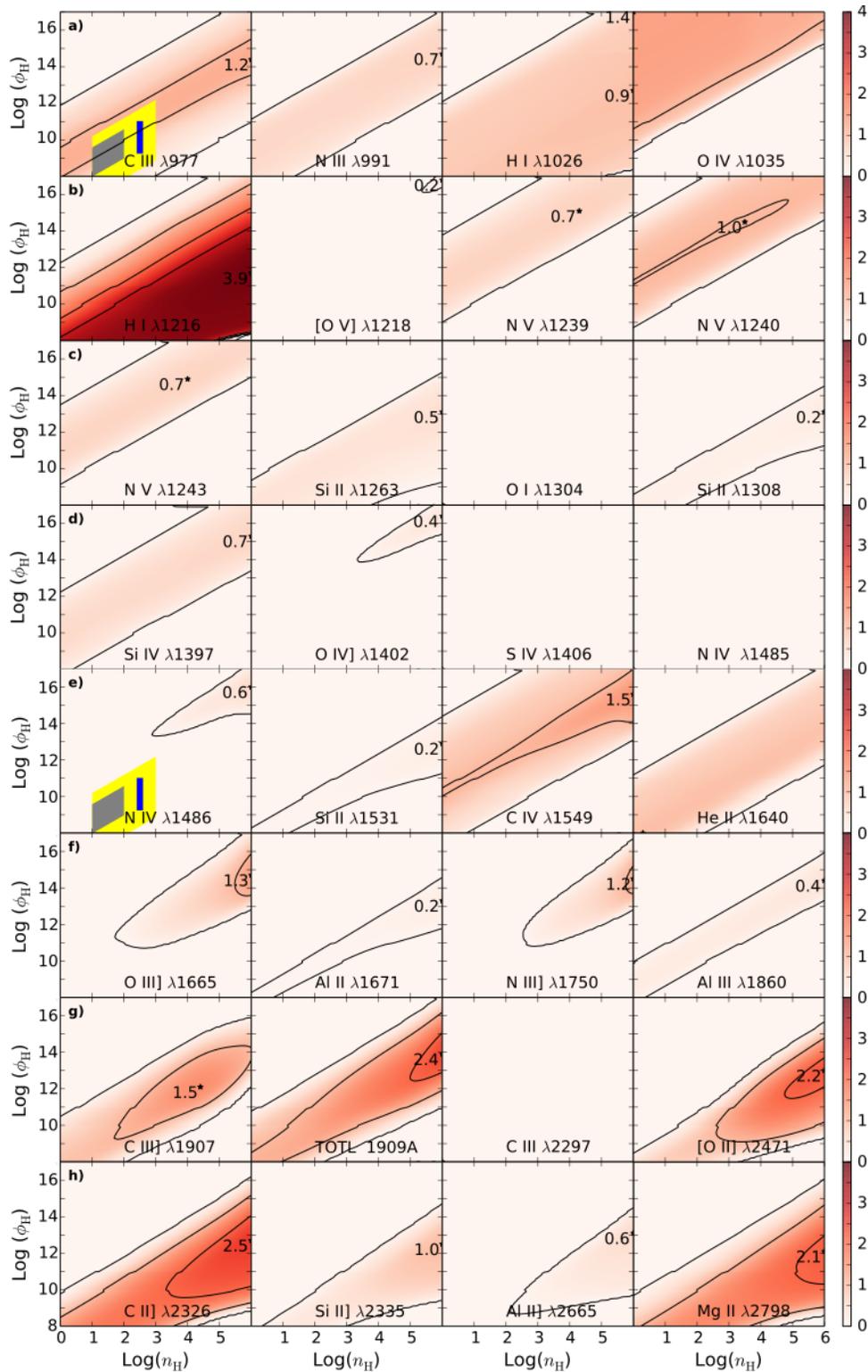

**Figure 4a.** Baseline model UV emission lines. Contours of log equivalent widths, relative to the continuum at 4860Å, for 32 UV emission lines as a function of hydrogen ionizing photon flux ($\phi_H$) and hydrogen density ($n_H$). The contour increments are 1 dex in black filled in with a color map with the labeled black star indicates the peak equivalent width.

*3.3.2 Optical Emission Lines*

Figure 4b displays the equivalent widths across the LOC plane for selected optical emission lines. Many of the $H^0$, $He^0$, and $He^+$ recombination lines are emitted over a wider area on the LOC plane than the other optical emission lines, including at low $\phi_H$ and high $n_H$ regions, because H I, He I, and He II have $n_{crit} \sim 10^{15}$ cm$^{-3}$ and thus are usually in LDL. At high $U$ values, however, the high $T_e$ causes the recombination coefficient to decrease making recombination less likely and causes large declines in $W_\lambda$ of the Balmer lines, He I λ5876, and He II λ4686. These two emission lines exhibit peaks at similarly high densities, but different $\phi_H$ values. Note also that our simulations do not predict particularly strong He II λ4686 emission. A strong He II λ4686 line is indicative of more He+ ionizing photons and simple photoionization models often under-predict the line in relation to the rest of the optical spectrum (Ferguson, Korista, & Baldwin 1997, Ferland & Osterbrock 1986). Our models do however predict weak [Ar IV] λ4711 emission (Figure 4b, row e), yet still strong enough to be detectable by current SDSS spectrographs and less sensitive spectrographs when co-adding similar spectra (Richardson et al. 2016).

BPT diagrams constructed with the ratios of [O III] λ5007 / Hβ and [N II] λ6584 / Hα have been useful in separating H II region galaxies from active galaxies. Our study supports recent work that has shown that selectively emphasizing different parts of the LOC plane gives different emission line ratios, which, in turn, give different results when using the BPT diagram (Richardson et al. 2016). Both [O III] λ5007 and [N II] λ6584 show emission on our grids from the bottom left along a constant ionization parameter (Figure 4b, rows f and h). [O III] λ5007 emits at higher $\phi_H$ and $n_H$ values, but also peaks at higher $U$. Their peak $W_\lambda$ are similar, only 0.2 dex different (2.9 for [O III] λ5007 and 2.7 for [N II] λ6584), but the peak $W_\lambda$ of [O III] λ5007 is located at a slightly higher $\phi_H$ value (13.2 and 10.5 respectively). Additionally, O [III] λ5007 peaks at $n_H$ = 4.8 whereas [N II] λ6584 peaks at $n_H$ = 3.9. Lastly, [N II] λ6584 emits along a broader range of ionization parameters than [O III] λ5007. Both Hα and Hβ emit along a broad range of ionization parameters (Figure 4b, rows h and f). The only regions in which they do not emit are the optically thin regions (upper left and lower right corners). It is thus clear that emission lines from metals, as well as many others, emit differently in different parts of our grid.

As with UV emission lines, there are various indicators of physical conditions among the optical emission lines. For example, the ratio of [O III] (λ4959 + λ5007) / λ4363 is an electron temperature indicator (Figure 4b, rows f and d). A smaller ratio indicates a higher electron temperature. As Richardson et al. (2014) note, at high densities, [O III] (λ4959 + λ5007) / λ4363 further decreases, reflecting mainly a drop in [O III] λ5007 due to collisional quenching and steady emission from [O III] λ4363. Consequently, at these high densities, this ratio does not serve as an accurate temperature indicator, however such high densities are not thought to occur in H II regions. Other such temperature indicators include [O I] (λ6300 + λ6364) / λ5577 (Figure 4b, rows g, h, and f). The ratios of various collisionally de-excited lines can provide an $n_H$ probe. Two examples of lines that can be used to determine $n_H$ are [O II] λ3729 / λ3726 (Figure 4b, rows b).

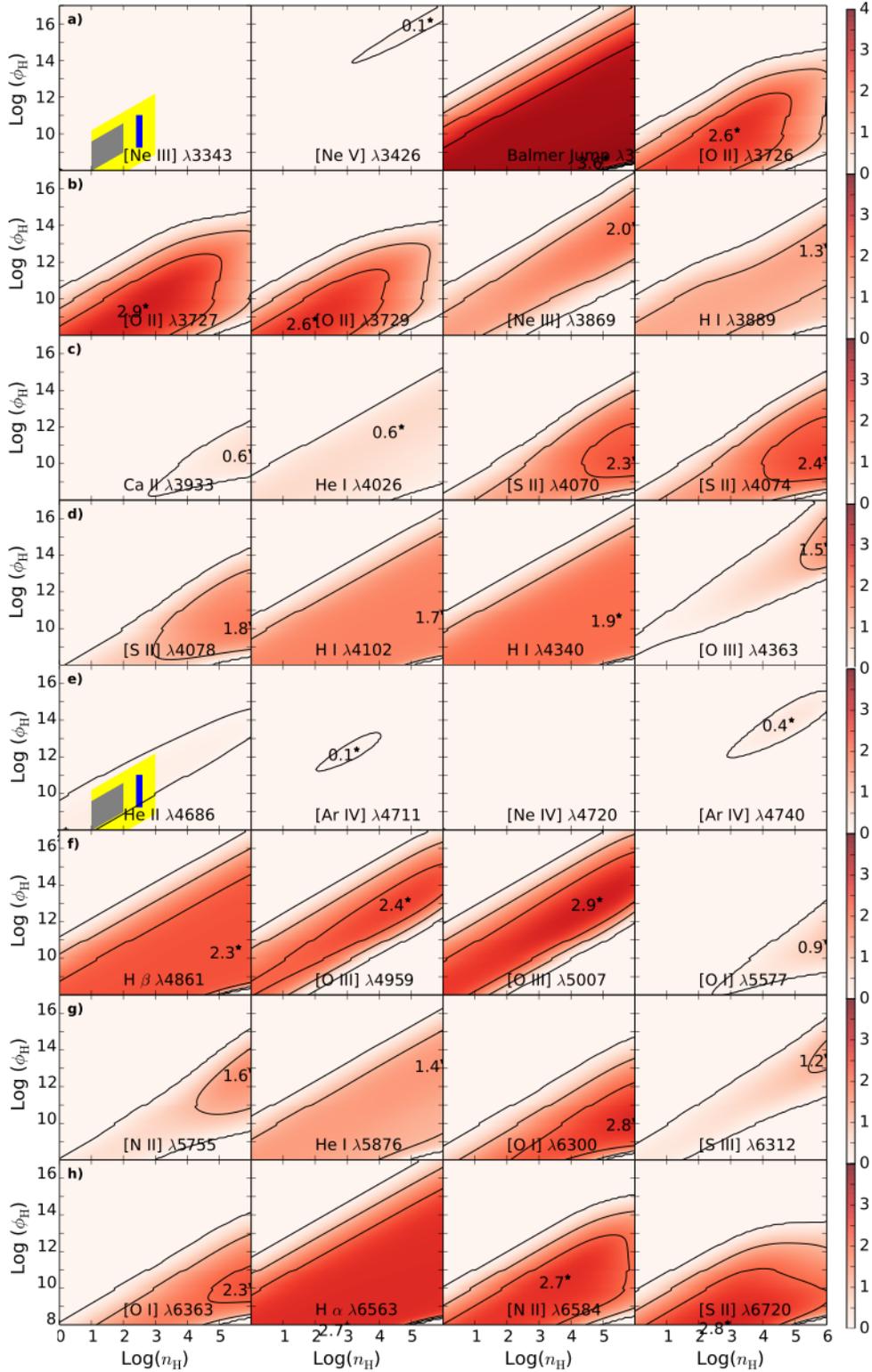

**Figure 4b**. Baseline model optical emission lines. Contours of log equivalent widths in the same manner as Figure 4a for optical emission lines.

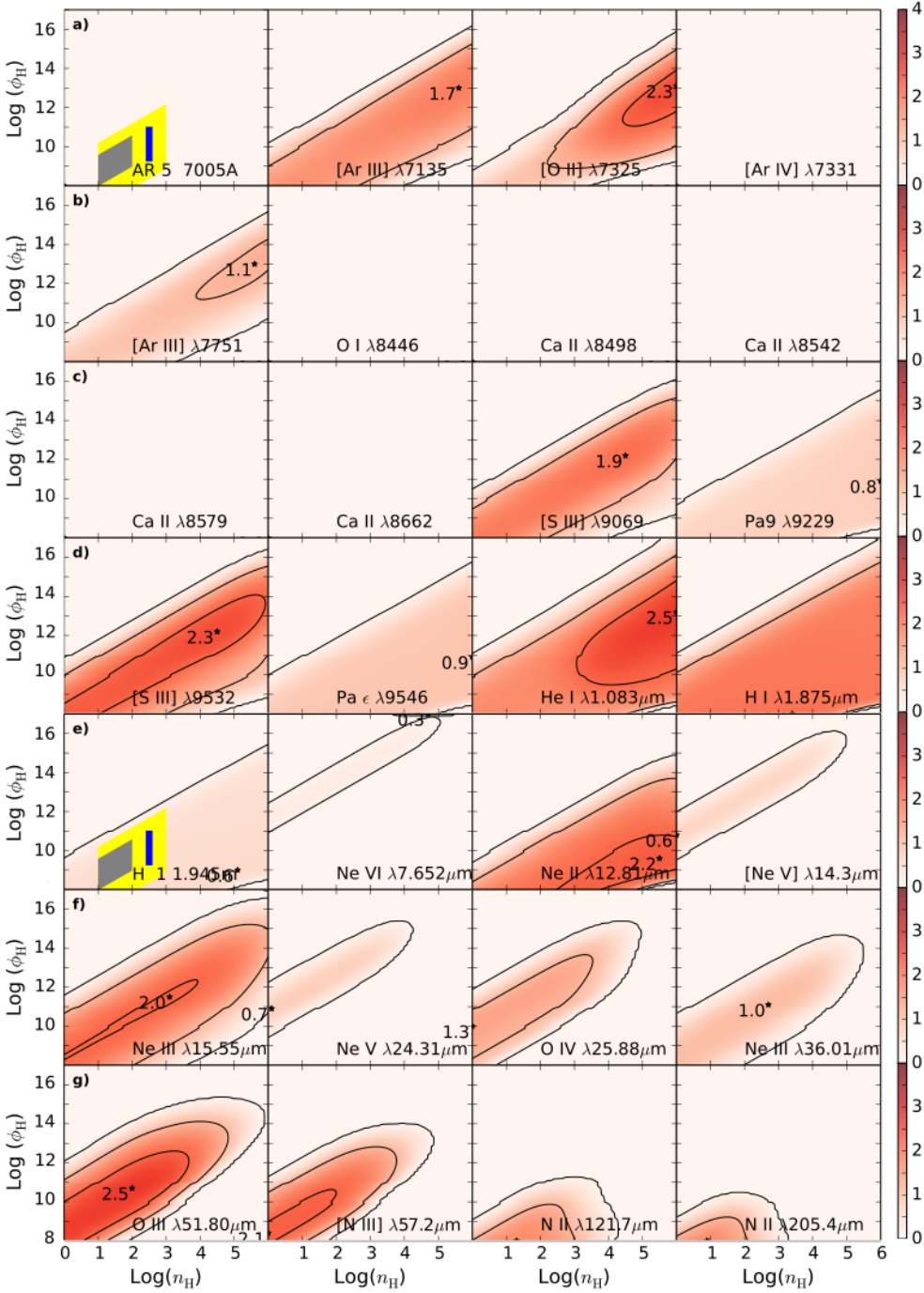

**Figure 4c.** Baseline model IR emission lines. Contours of log equivalent widths in the same manner as Figure 4a for IR and fine structure emission lines.

*3.3.3 IR Emission Lines*

Figure 4c displays the equivalent widths across the LOC plane for selected IR emission lines. Most of the infrared emission in our study is constrained to the bottom left of our grids, a parameter space that corresponds to low $n_H$ and low $\phi_H$ values. Although grains influence IR emission, grains in H II regions are not as important as in PDR regions where photoelectric heating serves as the dominant excitation source (AGN3). However, the IR emission lines that we track emit most efficiently in low $n_H$ and low $\phi_H$ regions and their emission cuts off conspicuously close to where we phase out grains. Because of this trend, we decided to compare our baseline dusty model with a dust-free model. We found that grains did not make a large difference in the peak equivalent widths, which means that they do not influence the strength of the IR emission lines that we are tracking. For example, for [N II] 122 µm, the peak $\log(W_\lambda)$ = 1.6 for the dusty case and $\log(W_\lambda)$ = 1.5 for the dust-free case and the peak $W_\lambda$ of [O III] 52 µm was only twice as high in the dust-free case than in the dusty case (Figure 4c, row g).

Given the modest effect of dust on these emission lines, the natural explanation is that most IR lines reach their critical densities when $\log(n_H) > 5$ and thus they do not emit efficiently in high density regions because they are collisionally suppressed. This is primarily because IR lines typically have lower Einstein coefficients, $A_{ij}$, compared to shorter wavelength emission lines, although exceptions do exist. For example, $\log(n_{crit}([N II] 122 µm)) = 2.56$ and it most efficiently emits around $\log(n_H) = 3$ (Figure 4c, row h). Similarly, the $\log(n_{crit}[O III] 52 µm) = 3.25$. This is a few orders of magnitude lower than its optical equivalent, [O III] λ5007, whose $\log(n_{crit}) = 6.43$ (Rubin 1989). Clearly, the [O III] 52 µm emission line still emits when $\log(n_H) > 3.4$ but the region it emits most efficiently is $\log(n_H) = 2$ (Figure 4c, row g; Rubin 1989).

Various IR fine-structure emission lines can also be used to predict electron temperatures with a given electron density. Such predictions can be made using ratio of [O III] 88 µm and [O III] λ5007. Though the [O III] λ5007 / [O III] 88 µm ratio depends on both temperature and density, the common [S II] λ6716 / [S II] λ6731 ratio can break its degeneracy. Consequently, by measuring both these ratios, we can determine the average values of both $T$ and $n_H$ (AGN3). Calculating these ratios from our grids indicates that our Cloudy simulations predict electron temperatures around $10^4$ K with $\log(n_H) \sim 3.0$, which is consistent with Figure 3. De Looze et al. (2014) also finds [O III] 88 µm shows a strong correlation with the SFR. We discuss the metallicity sensitivity of [O III] 88 µm further in the sensitivity studies section (§4.2).

Abel and Satyapal (2008) study [Ne V] emission in what they expect to be starburst galaxies, determining that it is almost always due to unobserved AGN activity. Our grids do predict some [Ne V] 14.3 µm and [Ne V] 24.3 µm emission; however, this emission is minimal, peaking at 0.6 dex and 0.7 dex respectively (Figure 4c, rows e and f) both at very low $n_H$. This seems to confirm their predications that starbursts produce little [Ne V], and strong [Ne V] emission is likely due to AGN activity or fast shocks, however the simple presence of [Ne V] emission should not attributed to non-thermal excitation. Observations of [Ne V] above 0.5 dex should serve as a red flag that other excitation sources must be at play beyond solely starlight.

**4. Sensitivity Studies**

In this section, we will discuss the sensitivity of our model to column density, gas metallicity, star-formation history, and dust. For our baseline model, we made assumptions about these values. Here, we explore the results of relaxing these assumptions.

## 4.1 Column Density

We begin by exploring the effects of relaxing the column density criteria. For our baseline model, the stopping condition is either when the simulation converges or when $N(H) = 10^{23}$ cm$^{-2}$ is reached. When the column density criteria is no longer supplied, Cloudy has difficulty converging upon a solution with the calculations $\log(\phi_H) > 21$. Because our simulation grid ranges from $8 < \log(\phi_H) < 23$, it is necessary to include the $N(H)$ stopping criteria. However, if we take the restriction off, we find, for the most part, that there is no significant difference in the strength of the emission lines with $\log(\phi_H) < 21$. However, these simulations are not able to capture many of the peak equivalent widths for emission lines that are peaking at high $\log(n_H)$ and high $\log(\phi_H)$ because Cloudy was unable to handle these conditions. We find that the column density stopping criteria is necessary for our simulations to capture many of the peak $W_\lambda$, and that not having this criteria may only affect emission line strengths in the most extreme conditions.

## 4.2 Metallicity

We have also explored the impacts of varying the metallicity from $Z = 0.2\ Z_\odot$ to $Z = 5.0\ Z_\odot$ in the cloud. Since varying both the $Z$ of the SED and of the cloud simultaneously would not allow us to interpret the effects of each independently, we chose to only study the effects of varying metallicity of the cloud. To adopt alternate metallicities for the cloud region, we linearly scale all of the metal number abundance by a scale factor $\xi$, the notable exception being nitrogen, which scales as $Z^2$ due to secondary nitrogen production when N is synthesized from C and O (Baldwin et al. 1991, K97). Once $\xi$ is input, we calculate the metallicity according to the following relation (Hamann et al. 2002):

$$\frac{Z}{Z_\odot} = \frac{\xi}{X_\odot + Y_\odot + (2\xi - 1)Z_\odot} \tag{4}$$

Assuming $\Delta Z = \Delta Y$ as determined by Baldwin et al. (1991) we can determine the hydrogen, helium, and metals abundances by mass fraction since $X + Y + Z = 1$. Calculating the helium number abundance scale factor is then straightforward and $\xi \approx Z/Z_\odot$ for small deviations from $Z_\odot$. For the subsolar case ($Z = 0.2\ Z_\odot$), we also add cosmic rays around $\log(n_H) = 7$ because the gas becomes partly molecular, which can contribute to excitation, however their inclusion had a negligible effect on emission line strengths.

We note that despite having adopted a different stopping condition to ensure that most of our simulations reach the ionization front (as described in §3.1.2), some emission lines still show a pocket of no emission in the bottom left of the LOC plane (e.g. C IV $\lambda$1549 in Figure 5a, row d and [O III] $\lambda$5007 in Figure 5b, row c). This pocket of no emission was neither present in our solar simulations nor in our subsolar simulations, but it was present at higher metallicity, where there are more metals to absorb the incident radiation field. The presence of more metals decreases the overall ionization of the gas and makes it difficult for high ionization potential emission lines to emit, like C IV $\lambda$1549, with few ionization photons available above 47.9 eV. Interestingly, we do not see this pocket of no emission for C III $\lambda$977 but we do this for [O III] $\lambda$5007, despite each line requiring photons with approximately the same ionization energy. This is because continuum pumping and recombination contribute to C III $\lambda$977 emission, whereas no such mechanisms contribute to the emission of [O III] $\lambda$5007.

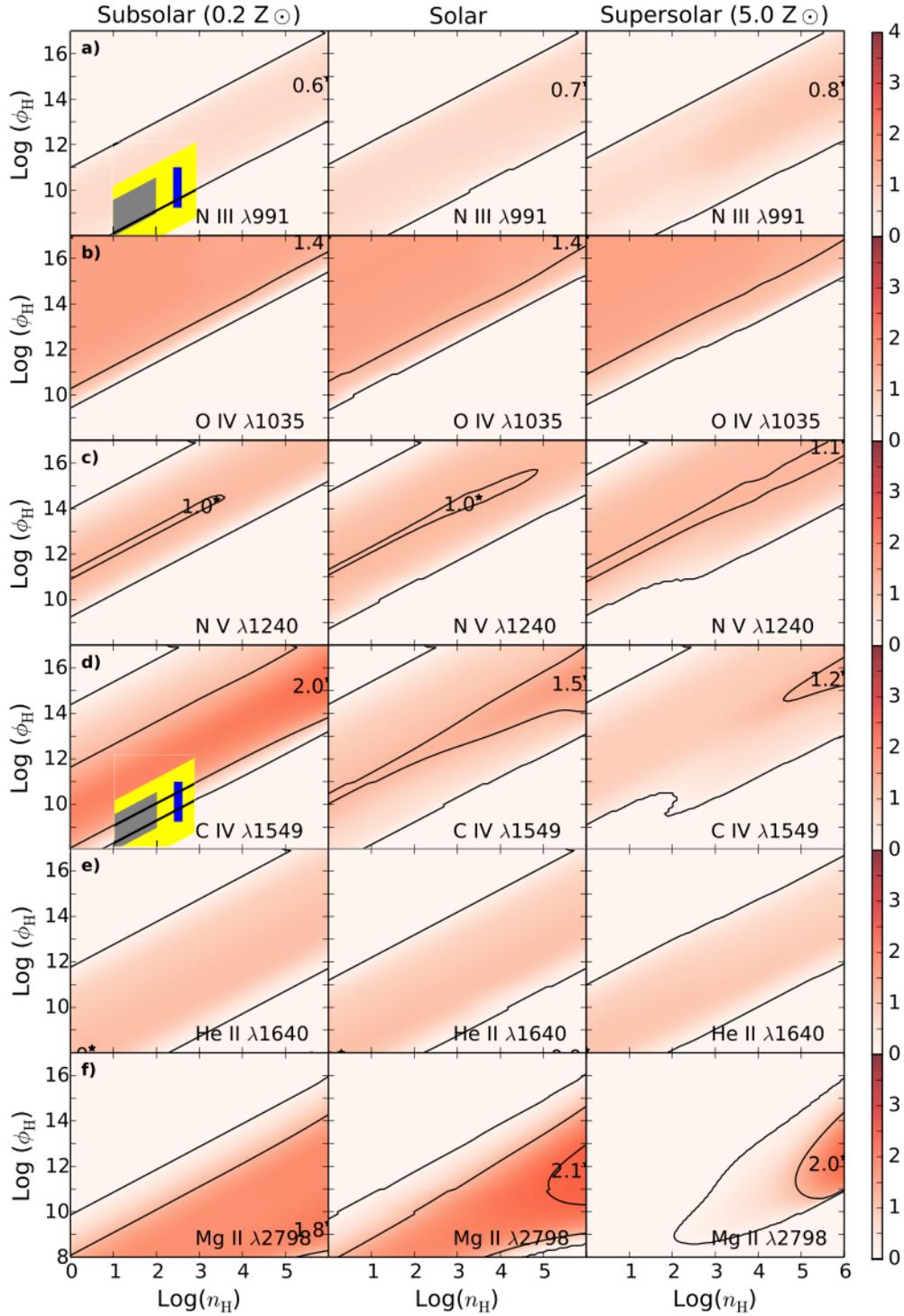

**Figure 5a**. Predicted UV emission as a function of metallicity. The left panel shows equivalent widths for 0.2 M$_\odot$, the middle panel shows solar metallicity, and the right panel shows equivalent widths for 5.0 M$_\odot$.

*4.2.1 UV Emission Lines*

Figure 5a displays the equivalent widths across the LOC plane for selected UV emission lines as function of metallicity. In general, we observe that with increasing metallicity, most of the UV emission lines increase in peak emission and some begin to emit over a wider range of ionization parameter (Figure 5a, rows a-c). We observe that a few UV emission lines, e.g. H I λ1216, O I λ1304, N IV λ1485, He II λ1640, and C IV λ1549, decrease in strength. Following Ferland et al. (1996) and K97, we will specifically discuss the relationships between oxygen, nitrogen, carbon, and helium.

Since we scale nitrogen with $Z^2$ due to secondary nitrogen production, our grids also show that N III λ991 increases 25 fold to $\log(W_{\text{N III}}) = 2.4$ at 5.0 $Z_\odot$ (Figure 5a, row a). Similarly, they show that N V λ1240 increases with increasing $Z$ but much less drastically than N III λ991: the strength of N V λ1240 at 5.0 $Z_\odot$ is only 13 times the strength of N V λ1240 at 0.2 $Z_\odot$ (Figure 5a, row c).

The effects of the increase in nitrogen abundance can be observed in the strengths of carbon and oxygen. As Ferland et al. (1996) note, the sum of N V λ1240 and C IV λ1549 emission remains fairly constant with metallicity changes because the two lines together dominate the cooling in the more ionized regions of the cloud. We find this consistent with our simulations since the sum hovers around 5.0 dex. However, as metallicity is increased, the nitrogen abundance exceeds the carbon abundance resulting in nitrogen carrying much of the total cooling. Since the cooling shifts from carbon and oxygen to nitrogen, the emission of C IV λ1549 is suppressed. On our grids, the peak $\log(W_{\text{C IV}})$ decreases 0.4 dex from 0.2 $Z_\odot$ to 5.0 $Z_\odot$ (Figure 5a, row d), while the peak $\log(W_{\text{N V}})$ emission increases 1.1 dex (Figure 5a, row c).

Additionally, as Ferland et al. (1996) discuss, He II λ1640 decreases with increasing $Z$ due to the increased abundance of heavy elements which contribute to an increasing fraction of the total gas opacity and absorb much of the helium-ionizing radiation. Accordingly, our grids show that at 5.0 $Z_\odot$ He II emission is approximately 0.6 the He II emission at 0.2 $Z_\odot$ (Figure 5a, row e).

*4.2.2 Optical Emission Lines*

Figure 5b displays the equivalent widths across the LOC plane for selected optical emission lines as function of metallicity. Many of the optical emission lines decrease in strength with increasing metallicity. For example, the [Ar IV] λ4740 emits at low metallicity (0.6 dex) but its emission decreases to zero at high metallicity (Figure 5b, row b). This general trend can be explained through the thermostat effect: though metal abundances increase when metallicity is increased, the amount of coolants also increases (especially in the case of nitrogen) and the cloud decreases in electron temperature. Emission line strengths are more strongly dependent on electron temperature than abundance, so the increase in coolants decreases the strength of the emission lines. In addition to decreasing in strength with increasing metallicity, the emission of our optical emission lines becomes more concentrated towards the center of the grids, along log $U = 0$.

As expected, due to our scaling of nitrogen, the nitrogen emission lines increase in strength. The peak equivalent width of [N II] λ6584 is nearly 40 times higher on the supersolar grids than the subsolar (Figure 5b, row f). We also see [O III] λ5007 decrease in strength with increasing metallicity (Figure 5b, row c). It should also be noted that ([O II] λ3727 + [O III] λλ4959, 5007)/Hβ acts as a metallicity indicator (Pagel et al. 1979, McGaugh 1991, Kewley and Ellison 2008). However, it does not give a unique solution because at low metallicities the ratio increases with increasing metallicity, and, at high metallicities, the ratio decreases because the cooling by the IR lines becomes more efficient. Thus, ([O II] + [O III])/Hβ should be analyzed considering additional metallicity-indicating line ratios (Nagao, Maiolino and Marconi, 2006, Raiter et al. 2010).

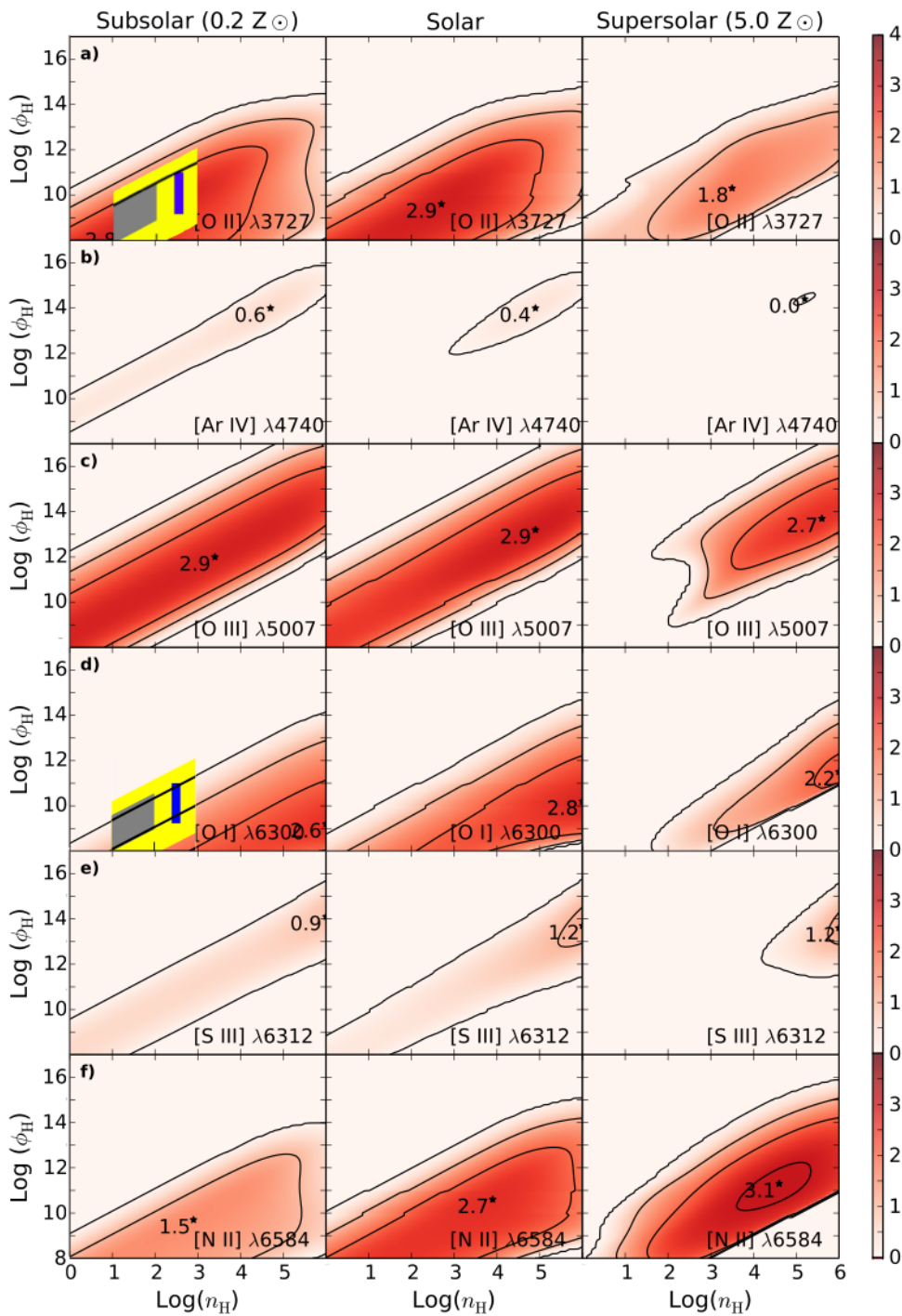

**Figure 5b.** Predicted optical emission as a function of metallicity. Contours of log equivalent widths for varying cloud metallicities in the same manner as Figure 5a for optical emission lines.

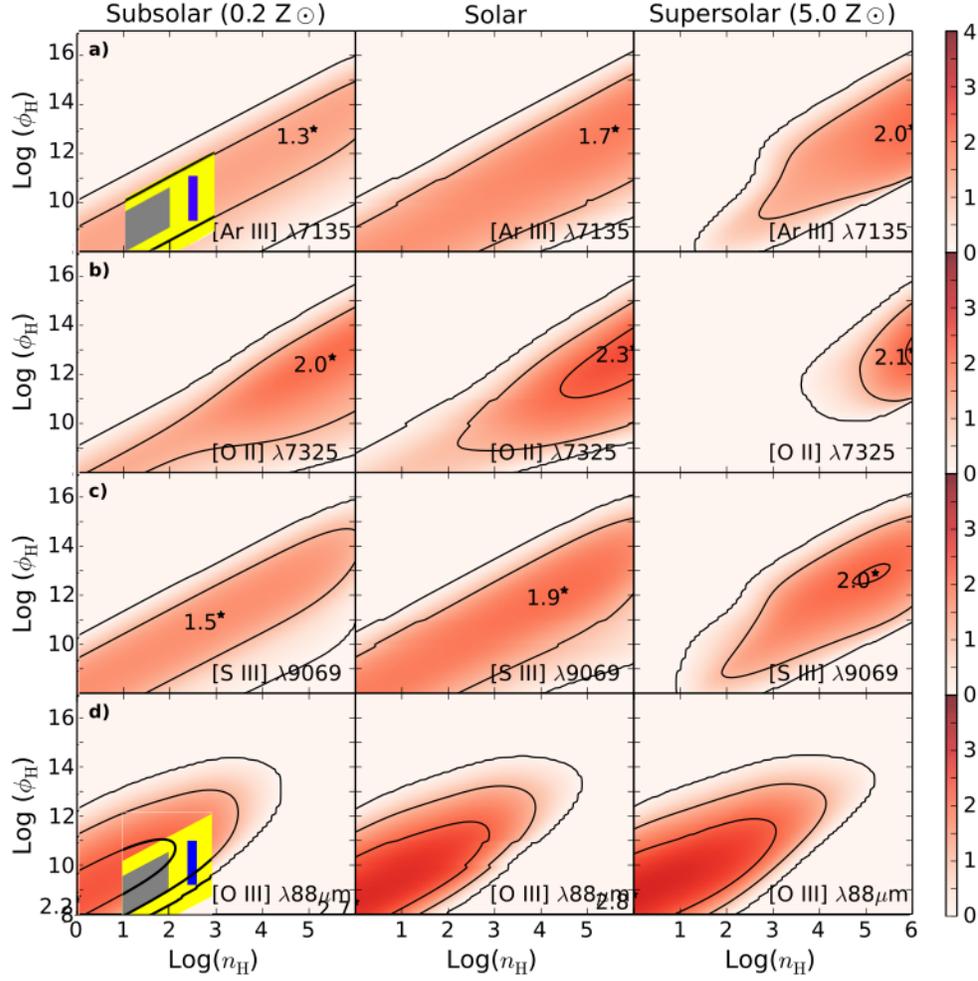

**Figure 5c.** Predicted IR emission as a function of metallicity. Contours of log equivalent widths for varying cloud metallicities in the same manner as Figure 5a for IR and fine structure emission lines.

*4.2.3 IR Emission Lines*

Figure 5c displays select IR emission lines' equivalent widths across the LOC as function of metallicity. IR emission line strengths generally increase with increasing metallicity. This is because when the electron temperature of the cloud is low, as in the case of high metallicity, the cooling is shifted from the UV and optical lines. As the metallicity continues to increase, the IR lines are able to act as more efficient coolants. As noted above, many of these lines cease to emit in the lower right corner of the LOC plane, but their peak emission increases. Consequently, [Ar III] λ7135 emission nearly quadrupled, [O II] λ7325 emission was over three times as strong, and [S III] λ9069 tripled with the higher metallicity simulation (Figure 5c, rows a, b, and c).

Our simulations show that the peak emission of [O III] 88 μm increased with increasing metallicity (Figure 5c, row d). As discussed earlier (§3.3.3), De Looze et al. (2014) show that [O III] 88 μm emission has an especially strong correlation with SFR. We have found a noticeable correlation with increasing metallicity, about a factor of four, and substantial variation across the LOC plane. Given that this emission line does show significant variation with $\phi_H$, $n_H$, and Z, observers should be cautious about using it a SFR indicator when the physical conditions are widely varying.

## 4.3 Star-formation History

We previously discussed the spectral energy distribution we have adopted, however here we explore the effects of varying the SFH on the peak equivalent width predictions. Figures 6a-c show the effects of adopting continuous and instantaneous Padova tracks and continuous and instantaneous Geneva rotation tracks on select emission lines from the UV to the IR at ages 0 Myr, 2 Myr, 4 Myr, 5 Myr, 6 Myr, 8 Myr, In these figures, the peak equivalent widths of each emission line are tracked with age. It is worth noting that the peaks of the emission line presented may occur at different $\phi_H$ and $n_H$ values with different ages. Though this information is contained in the LOC plane, it is not presented as part of Figures 6a-c.

Nearly all the peak equivalent widths of the emission lines we track decrease with time when we adopt any of the four evolutionary tracks. This is unsurprising considering the general decrease of high-energy photons with later ages (§3.1.1) Since the model atmospheres for the continuous star formation show little change in spectral slope as a function of cluster age, the continuous star formation models give similar results to zero-age instantaneous models. However, the instantaneous models, as evident in Figure 2, give few high-energy photons at ages greater than 6 Myr, and consequently, emission lines' peaks decrease.

When comparing only the two continuous evolution tracks, there is little observable difference. High ionization emission lines are the main exceptions to this trend. For example, [Ne V] λ3426 only emits when the Padova tracks are adopted; specifically, [Ne V] λ3426 emission dies off after 5 Myr with the Padova instantaneous track but continuous to emit past 5 Myr with the Padova continuous track (Figure 6b).

We observe that most emission lines die off after 5 – 8 Myr when adopting either of the instantaneous tracks. The Geneva instantaneous track tends to produce stronger emission than the Padova instantaneous track, likely due to its incorporation of rotation. Nonetheless, the Padova instantaneous track produces stronger high ionization emission lines. For example, the optical lines [O I] λ5577, [N II] λ5755, [O III] λ5007, and [S II] λ6720 are all stronger with the Geneva track, while [N V] λ3426 is stronger with the Padova continuous track (Figure 6b).

Dust obscuration makes the first few Myr after stellar birth inaccessible to detailed age-dating; however, we know that in these first few million years, O-type stars tend to dominate the luminosity of starburst galaxies. In our simulations, there is not much observable difference in emission lines' peak equivalent widths between the first few Myr for different evolutionary tracks since all of our tracks start similarly. While most emission lines strengths remain constant, [Ar IV] λ4740 decreases in emission by a factor of 4 over this period of time (Figure 6b).

As the hot, young starburst ages to 4-5 Myr, stellar wind lines (e.g. C IV λ1550 and Si IV λ1400) dominate the emission in the wavelength region from 1200 to 2000 Å. These also include UV carbon and oxygen emission lines (Schaerer 2000). Generally, the optical and IR region lack features from the stellar atmospheres but the UV emission lines tend to remain strong. In our simulations of the Padova instantaneous track, the UV emission lines decrease on the order of 0.5-1.0 dex from 4-6 Myr (Figure 6a). The optical and IR line emission (for the same SFH) decrease on the order of 1.0-1.5 dex (Figures 6b and c). Notably, He II 4686, and [Ar IV] 4740 change significantly, ranging from a increase of 0.75 dex to an increase of 0.4 dex, respectively, between 3 and 5 Myr. The Padova and Geneva continuous tracks, however, do not show much difference between bands of emission lines through age.

The equivalent widths of many of the strong hydrogen recombination lines like Hα, Hβ, or Brγ can be used as age indicators because they measure the ratio of the young, ionizing over the old, non-ionizing stellar population. Our simulations fit this trend since our Hα and Hβ emission decreases about an order of magnitude with both instantaneous evolution tracks (Figure 6b). The effect of age is much more pronounced with the Padova instantaneous evolution track than with the Geneva.

After 5 Myr, the most massive stars in the starburst cool off and form Red Super Giants (RSGs). At 8 Myr, these RSGs dominate the near-IR portion of the stellar spectrum. When adopting the Geneva instantaneous track, emission line strengths begin falling off rapidly beyond 6 Myr (approximately 0.5 – 1.0 dex lower at 8 Myr than 6 Myr), especially in the case of the optical, most IR, and IR fine structure lines (Figure 6b and 6c). When adopting the Geneva and Padova continuous tracks, however, the emission line strengths remain constant across the 6-8 Myr range.

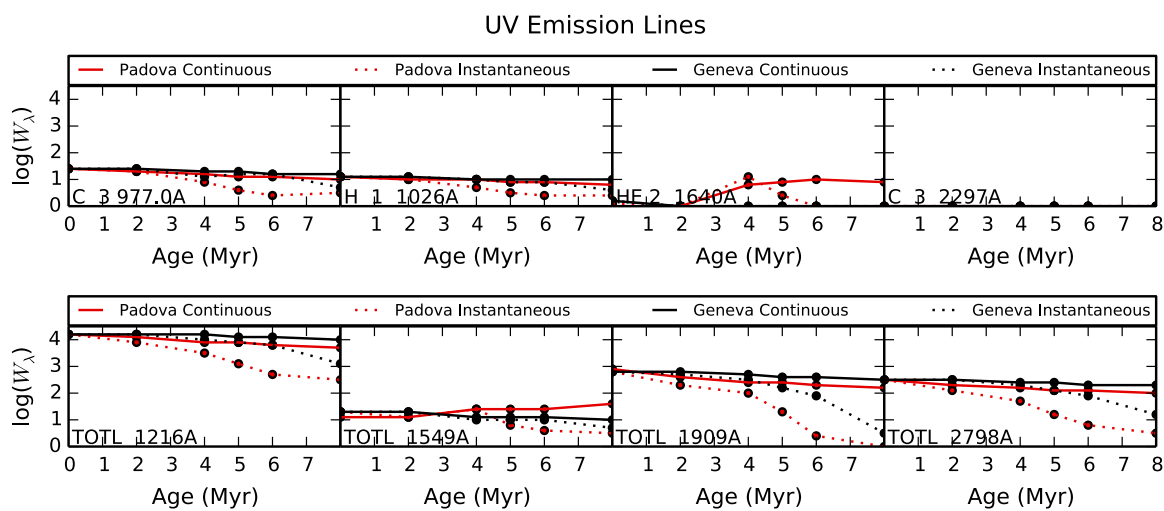

**Figure 6a**. UV emission line SFH comparison. SFH comparison for four evolutionary tracks (Padova continuous in solid red, Padova instantaneous in dashed red, Geneva continuous in solid black, and Geneva instantaneous in dashed black) tracked through age (0 to 8 Myr in 2 Myr intervals). This panel shows the peak $W_\lambda$ of select strongly emitting UV emission lines.

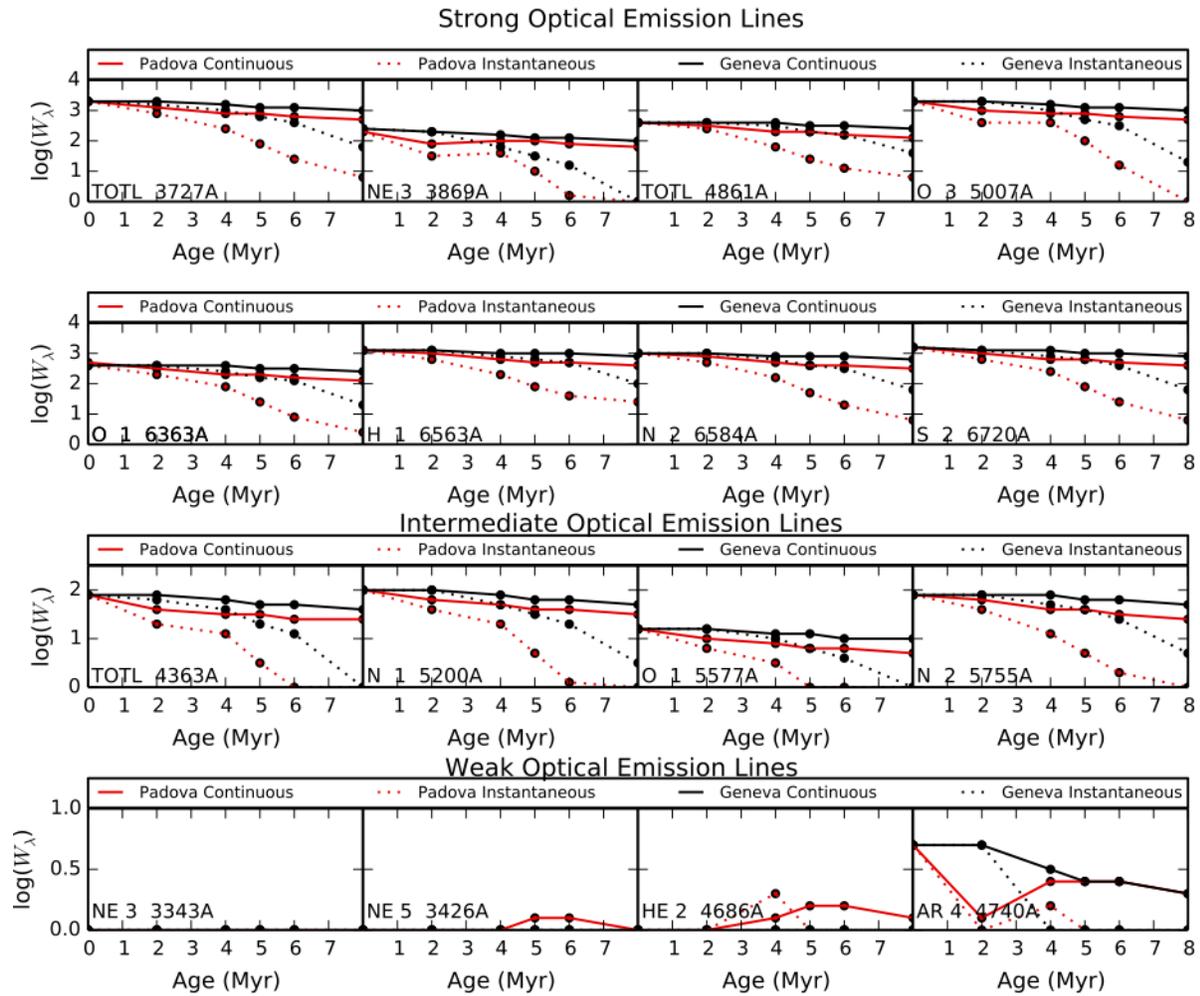

**Figure 6b**. Optical emission line SFH comparison. SFH comparison in the same manner as Figure 6a for optical emission lines.

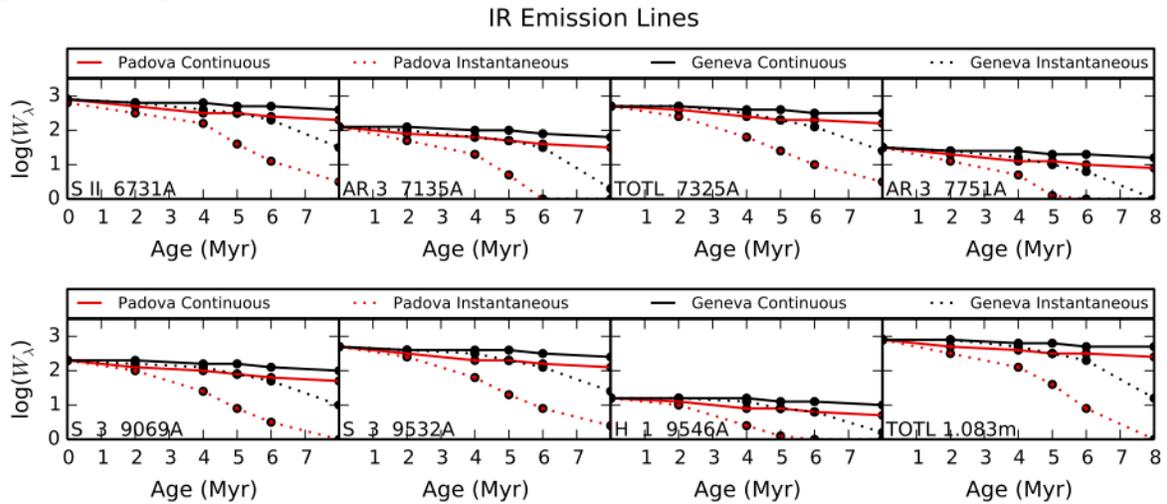

**Figure 6c**. IR emission line SFH comparison. SFH comparison in the same manner as Figure 6a for IR emission lines.

**4.4 Dust**

Though our baseline model includes grains, we have not yet analyzed the sensitivity of our LOC model to dust. Figures 7a - c displays the equivalent widths across the LOC plane for selected UV, optical, and IR emission lines comparing our baseline model to an entirely dust-free model with abundances given in Table 1.

Most of the emission lines we track maintain their shape across the LOC plane, with the range of ionization parameters over which they emit broadening slightly with the removal of dust. Generally, the effects of dust are most prominent with the UV emission lines and some of the shorter wavelength optical emission lines. This observation is consistent with other studies about the effects of dust on the UV emission lines coming from the gas clouds within starburst galaxies (e.g. Heckman et al 1998).

Overall, when comparing the dusty and dust-free simulations, we find the electron temperature across the LOC plane higher when dust is included. Ionized hydrogen and dust grains contribute equally to the heating of the cloud. However, the dust-free simulations have more coolants, since metals are not locked up in grains, making the overall electron temperature decrease. Due to the thermostat effect and the absence of photoelectric heating, this would typically lead to a decrease in metal emission line strengths (Shields and Kennicutt 1995); however, [Ne V] $\lambda 3426$, and [Ar IV] $\lambda 4740$ all show greater emission with the removal of dust (Figures 7a-c). The physical reason for this apparent contradiction is that dust makes a substantial contribution to the overall opacity in our dusty simulations, which decreases the availability of high-energy photons to ionize and excite the gas.

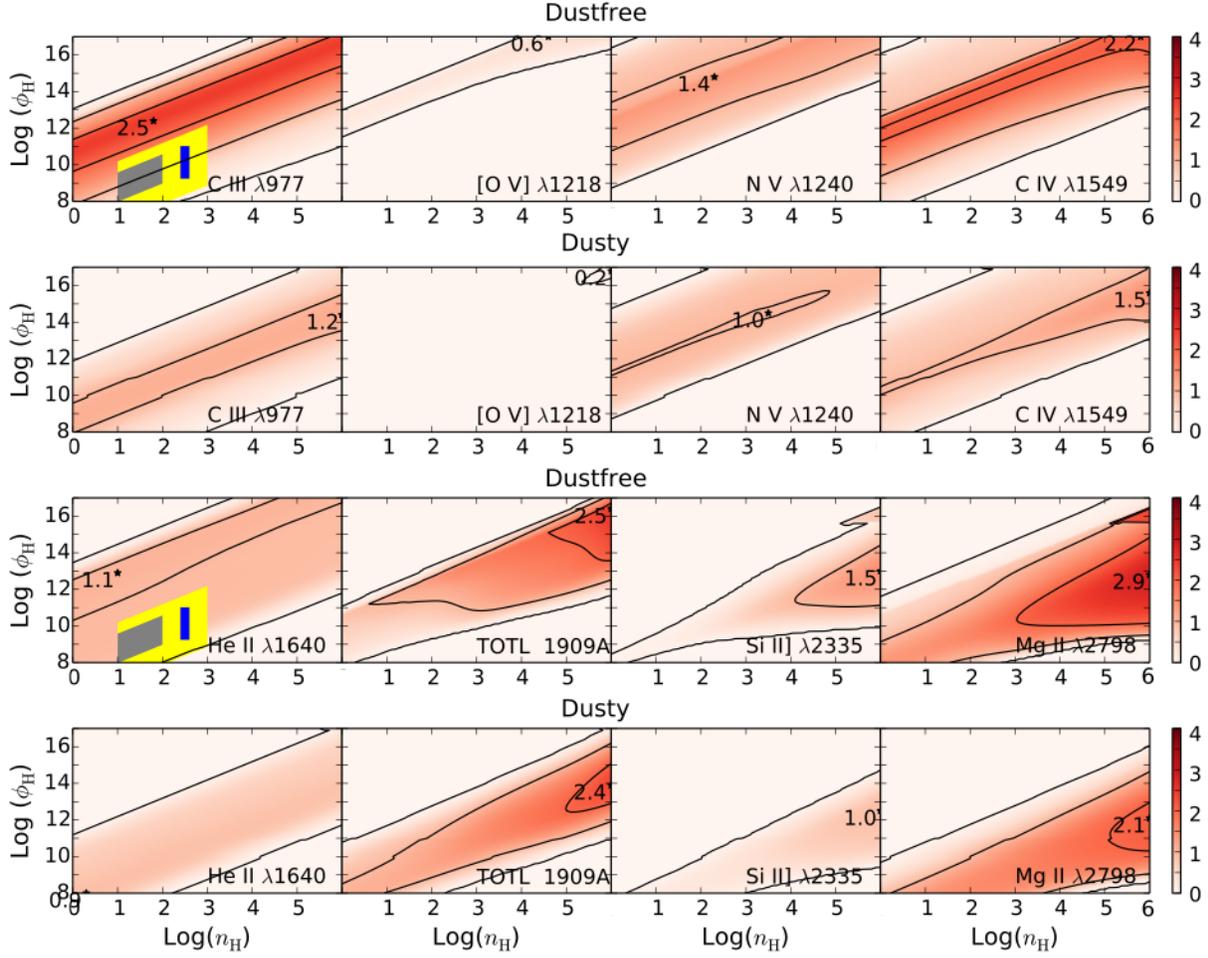

**Figure 7a**. UV emission lines dust comparison. Contours of log equivalent widths comparing our dusty and dust-free models for UV emission lines. Note that we phase out dust grains when $\log(\phi_H) > 17$ since dust sublimation occurs so these comparisons have been cut at $\log(\phi_H) = 17$.

Many of the equivalent widths of UV emission lines increase with the removal of dust since dust absorption peaks in the UV. Specifically, with the removal of grains, the peak equivalent width of N V λ1240 increases 0.4 dex, C IV λ1549 increases 0.7 dex, He II λ1640 increases 0.2 dex, and Si II] λ2335 increases 0.5 dex (Figure 7a, columns c, d, and a). One of the most drastic changes among the UV emission lines is shown by [O V] λ1218, which increases 0.4 dex with the removal of dust, while the region it emits across the LOC plane expands significantly (Figure 7a, column b).

Overall, several detached islands of emission appear or disappear with the inclusion or exclusion of dust. This effect is best seen with [S II] λ4078 and [Ar IV] λ4740 (Figure 7b, columns c and d). The most drastic change in the optical emission lines is shown by [Ne V] λ3426 which increases 0.6 dex with the removal of dust and [Ar IV] λ4740 which increases 0.8 dex with dust removal (Figure 7b, columns a and d). [O II] λ3727 decreases 0.4 dex with dust removal, while [O III] λ5007 increases 0.4 dex with dust removal (Figure 7b, columns b and a).

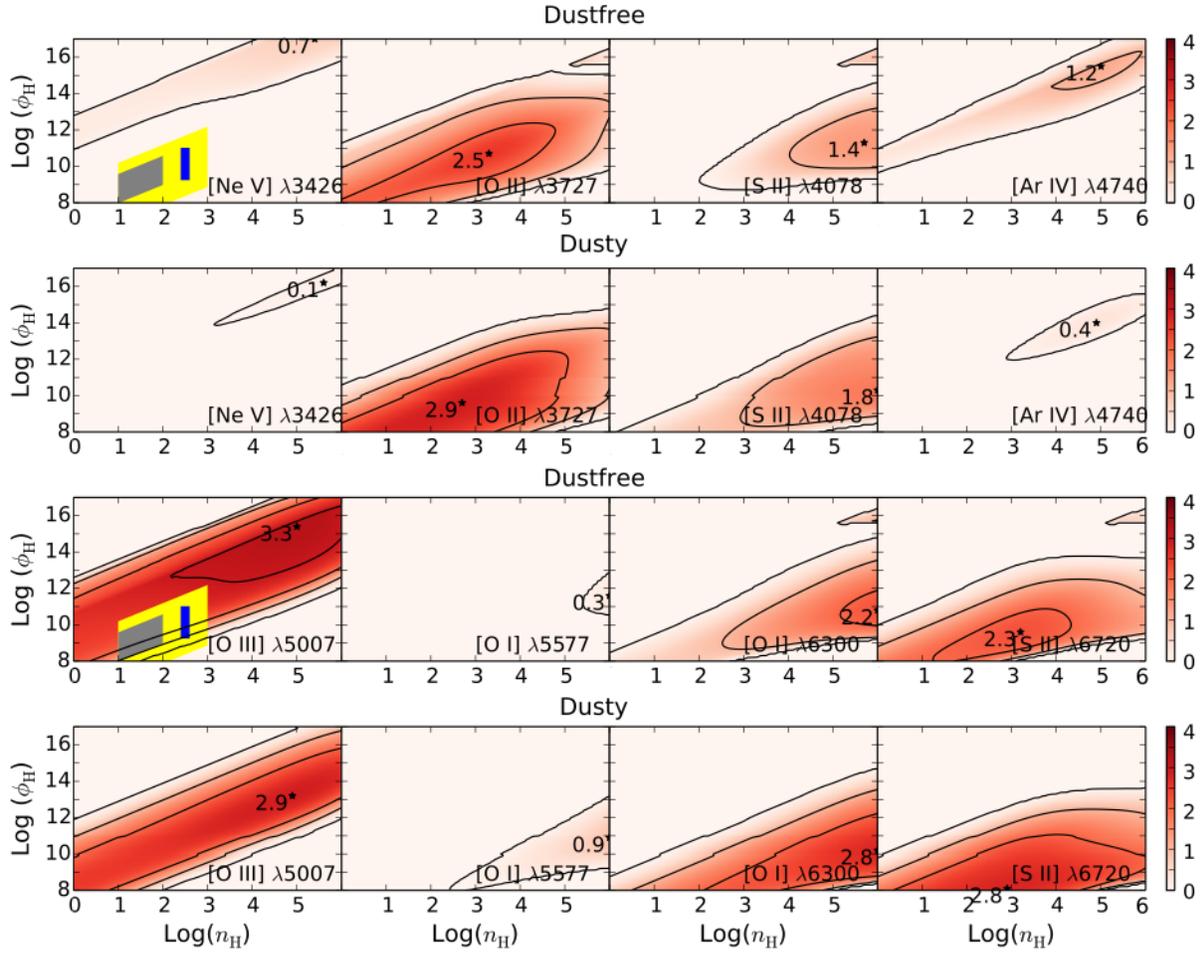

**Figure 7b**. Optical emission lines dust comparison. Dust comparison in the same manner as Figure 7a for optical emission lines.

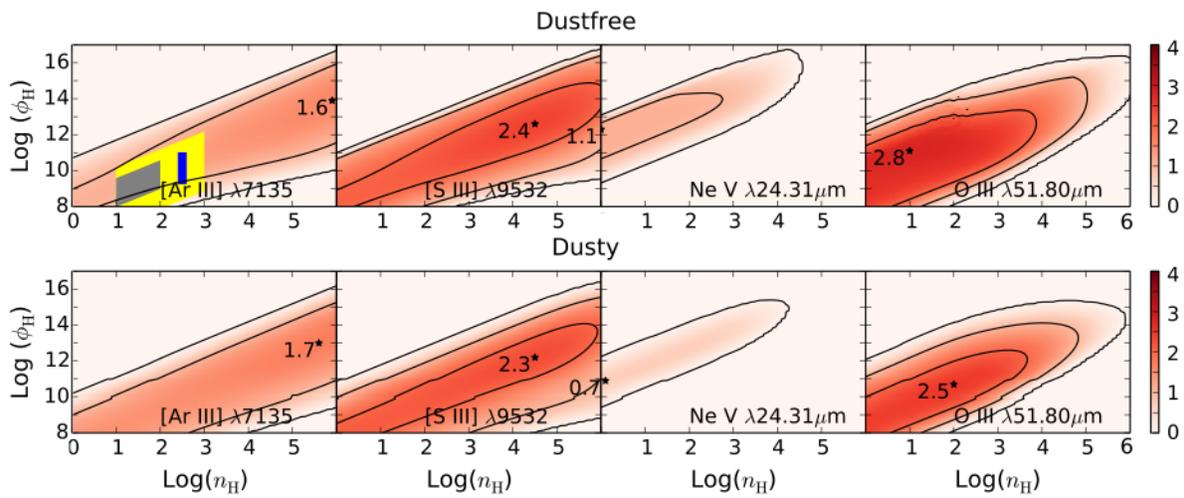

**Figure 7c**. IR emission lines dust comparison. Dust comparison in the same manner as Figure 7a for IR emission lines.

There is very little change shown by any of the IR emission lines. [Ne V] 24.31 µm changes the most, its peak equivalent width increasing 0.4 dex with dust removal, and [O III] 55.8 µm in a close second, increasing 0.3 dex with dust removal (Figure 7c, columns c and d). Otherwise, many of the peak log($W_\lambda$) of our IR emission lines and specifically, IR fine structure emission lines change by less than 0.2 dex.

## 5. Analysis

We begin by discussing the implications of our atlas on the local, low redshift galaxy literature presented in the introduction. We then move to discussing implications for future observations with JWST at higher redshifts. For specific comparisons of equivalent width predictions to observations, refer to Table 2 where we list both the peak equivalent width over the LOC plane and the equivalent width in gas conditions similar to local H II regions with log($U$) = -2.0 and log($n_H$) = 1.0.

As discussed in Satyapal et al. (2007), NGC 3621, an optically classified star-forming galaxy at low redshift, emits [Ne V] 14 µm and 24 µm. In our simulations, [Ne V] 14 µm and 24 µm get stronger with increasing metallicity. At solar metallicity, the peak log($W_{[Ne\ V]}$) of these emission lines are 0.6 and 0.7 respectively, and in high metallicity (5.0 $Z_\odot$) simulations they increase roughly by a factor of 2.5, to 1.0 and 1.1, respectively. Additionally, the emission for each of these lines increases similarly without dust grains. We note, however, that [Ne V] 14 µm and 24 µm emission on our grids begin at $U \approx 1.0$, a higher ionization parameter than what is typically observed locally (-3 < log($U$) < -1.5; Levesque et al. 2010). Thus, we agree with Abel & Satyapal (2008) and Guesva et al. (2000) that an additional excitation source, most likely an AGN, is needed for local [Ne V] 14 µm and 24 µm emission of this strength.

Lutz et al. (1998) report observations of local star-forming galaxies that show weak nebular [O IV] 25.9 µm emission without any signs of AGN activity. In our dusty, 5.0 $Z_\odot$ simulations, we find the peak log($W_{[O\ IV]}$) = 1.6, which is twice as strong as our baseline model emission and nearly 8 times as strong as our low-metallicity model; however, we again note that this [O IV] 25.9 µm emission only occurs when adopting the physical conditions atypical of the local star forming galaxies.

Lastly, Shirazi and Brinchmann (2012) report a significant number of optically classified star-forming galaxies with strong He II λ4686 emission around $z \sim$ 0-0.4 and were able to recreate this emission assuming 0.2 $Z_\odot$ in their simulations. We find that peak He II λ4686 emission does not change significantly as we vary from 0.2 to 5.0 $Z_\odot$ and that this emission does not occur in the range of local galaxies. Nonetheless, we do see minimal (log($W_{He\ II}$) = 0.2) emission in our 0.2 $Z_\odot$ simulations around log($U$) < -2 and low density. Thus, we suggest that perhaps there are low metallicity pockets within these local galaxies, which contribute to their overall He II λ4686 emission.

The *James Webb Space Telescope* (JWST), scheduled to launch in 2018, is ideal for IR observations of galaxies at higher redshifts than we have considered up until this point. Given the large influence of vigorous star formation on emission line production at early times in the universe (Madau & Dickinson 2014), JWST will be an ideal instrument to study these star-forming galaxies. JWST's instruments work in the range of 0.6 – 28 µm and will conduct deep-wide surveys of galaxies of $1 \leq z \leq 6$ in the rest-frame optical and near infrared. [1]

---

[1] http://www.stsci.edu/jwst/

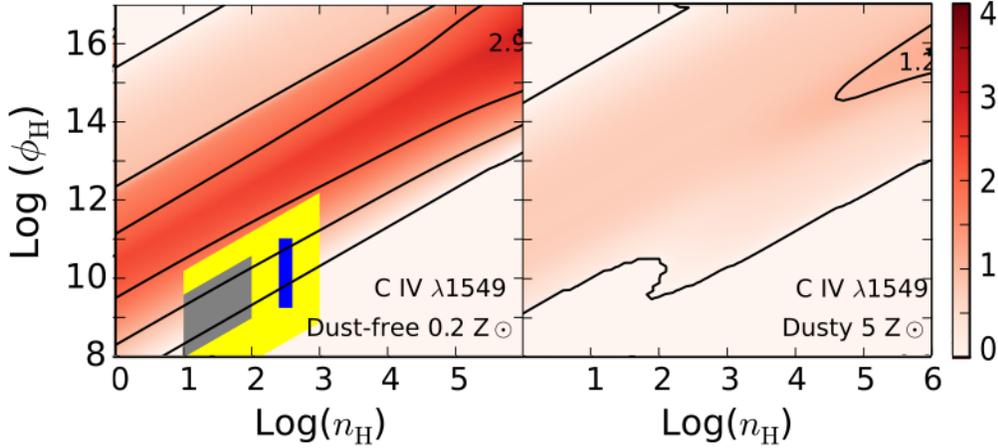

**Figure 8.** Comparison of C IV λ1549 for typical conditions in the local and higher redshift universe. The left panel is our dust-free, subsolar simulation (0.2 M$_\odot$) and the right panel is our dusty, supersolar simulation (5 M$_\odot$). C IV λ1549 should serve as a useful diagnostic of vigorous star formation given the wavelength band of *JWST* instruments because it emits most strongly under conditions unlike those typical found in nearby galaxies (low gas-to-dust ratio, low $Z$).

The emission lines O III] λ1661 and λ1666 get shifted to 0.6 – 0.8 μm range around $z \sim$ 3.0, while He II λ1640 is shifted to around 0.65 μm (Shapley et al. 2003), which are within 0.6 – 5.0 μm range of the JWST NIRCam. Further, optical and IR emission lines become accessible to the NIRCam in the $2 \leq z \leq 5$ range, for example, He II λ4686, [O III] λ5007, and [N II] λ6584.. We also expect that some IR emission lines, like [Ar III] λ7135 and [O II] λ7325 will be observable in this range with peak log($W_\lambda$) around 2-3.

With higher-$z$ galaxies, many of the UV emission lines get shifted into the range of JWST. For example, the N IV] λ1486 emitter in Raiter et al. (2010) at $z \sim$ 5.6 gets shifted into the range of NIRCam while the MIRI instrument on JWST is sensitive in the 0.5 to 28.3 μm and will provide medium resolution spectroscopy (R~3000) over this range.

Many of our UV emission lines accessible to JWST's MIRI well into the high-$z$ range are stronger in dust-free simulations than those of the baseline model and could thus be used as dust diagnostics (Figure 7a). For example, N V λ1240 and Si II] λ2335 increases 0.4 dex and 0.5 respectively when dust is not included. [O V] λ1218 both emits more strongly and in a greater range on the LOC plane without dust.

We predict that C IV λ1549 will be the most useful indicator of vigorous star formation at higher redshift with JWST observations. Given moderate ionization potential (47.9 eV respectively), the collisionally excited UV line C IV λ1549 should easily be formed and has been used in AGN literature to discriminate between pure shock and pure photoionization models of excitation (Allen, Dopita & Tsvetanov 1998). In the conditions of the early universe, we expect there to be less dust, low metallicity, and little AGN contribution since there are fewer supernova remnants, less chemical enrichment, and we are past the AGN epoch (z < 3) of galaxy evolution. C III λ977 and C IV λ1549 become stronger under these conditions (Figure 8).

Additionally, when adopting local nebular conditions, C IV λ1549 is not a strong line; therefore, it should only be detectable for higher redshift galaxies with little dust and low metallicity. The peak log($W_\lambda$) in our dusty 5 $Z_\odot$ simulations is 1.4 respectively while the peak

log($W_\lambda$) in our dust-free 0.2 $Z_\odot$ simulations is 3.0 respectively (Figure 8). It has strong emission at $8 \leq \phi_H \leq 12$, and $0 \leq n_H \leq 4$ in the dust-free case (1.0 < log($W_\lambda$) < 2.0). Given the sensitivity of JWST's MIRI instrument, we predict that these emission lines should be easily detectable in starburst galaxies at greater redshifts.

## 6. Conclusion

In this paper, we have compiled an atlas of predicted star-forming galaxy equivalent widths to be used by observers to constrain the physical conditions in the systems they observe. We sought to determine the extent to which we could reproduce high ionization emission lines seen by observers by only invoking photoionization via starlight; we do not employ photoionization by AGN or collisional ionization by shocks/turbulence. We thus began by asking what physical conditions are necessary in star-forming galaxies to produce these high ionization emission lines.

To address this question, we adopted a two-part methodology of simulating the star-forming region SED and then using a locally-optimally emitting cloud (LOC) methodology to investigate emission lines. Using Starburst99, we explored the parameter space of SFH and metallicity to determine what conditions would give the hardest spectrum. We considered the Geneva rotation tracks but found that the Padova AGB track SED with a continuous SFR produced the hardest ionizing spectrum. As we were investigating high ionization emission lines, we adopted this model as our baseline model. To account for the dust ubiquitous throughout H II regions, we consider dusty conditions while adopting a step function across the plane to account for sublimation. Finally, we adopt an LOC plane spanning $0 \leq \log(n_H) \leq 10$ and $8 \leq \log(\phi_H) \leq 22$ to match a wide range of physical conditions.

Having adopted the Padova AGB track SED at 5 Myr for our baseline model, we tracked 96 emission lines across the LOC plane. We found that collisionally excited UV emission lines reprocessed the spectrum along constant ionization parameter lines on the LOC plane. Many of our optical recombination lines emitted along a wider range of ionization parameter lines. We found that many of the optical emission lines that we tracked also exhibited an interesting double peak feature due to an ionization jump experienced by an element. This feature was even more evident in higher metallicity and dust-free simulations. We found that IR emission lines emit most efficiently in low $n_H$ and low $\phi_H$ regions on account of their low critical densities.

We next analyzed our model's sensitivity to metallicity, SFH, and dust. Nearly all the peak equivalent widths of the emission lines we track decrease with time when we adopt any of the four evolutionary tracks. There was little observable difference between continuous evolution models (except with high ionization emission lines of interest) and most emission lines die off after 5-8 Myr with the instantaneous models of evolution. We note that most of our emission lines maintain their shape across the LOC plane with a dust-free model, only changing slightly in their range of emission and peak log($W_\lambda$). Lastly, dust effects are most noticeable with UV emission lines and some of the lower wavelength optical emission lines.

UV emission lines generally decreased slightly with age and increased with dust removal (as dust absorption peaks in the UV). Most UV emission lines also increased in emission with increasing metallicity. Optical emission lines decreased in emission with increasing metallicity, decreased slightly with age, and were not particularly sensitive to dust. IR emission lines increase in emission with increasing metallicity, decrease slightly with age, and showed very little change with the introduction of dust.

In the end, we find that our grids suggest a pocket of more extreme conditions or AGN

activity when strong high ionization emission lines are present in the local universe. As we move to simulations with physical conditions more indicative of those found at higher redshift, we find our grids better at reproducing high ionization emission lines. In particular, the C IV λ1549 line should prove particularly strong at earlier times, making it an excellent candidate for indicating vigorous star formation in future JWST observations.

**Acknowledgments**

HM and CTR would like to thank the Elon University Lumen Prize, Honors Program, Summer Undergraduate Research Experience (SURE), and Student Undergraduate Research Forum (SURF). CTR would like to acknowledge Elon University for Reassigned Time and Summer Research Fellowships and acknowledge the support of the Extreme Science and Engineering Discovery Environment (XSEDE), which is supported by National Science Foundation grant number OCI-1053575. We greatly appreciate the insightful comments of Tony Crider, which improved the quality of this paper.

**Appendix A. Complete list of tracked emission lines.**

| Emission line | Wavelength | Notes |
|---|---|---|
| C III | 977 Å | |
| N III | 991 Å | |
| H I | 1026 Å | |
| O IV | 1035 Å | |
| Incident | 1215 Å | |
| H I | 1216 Å | |
| [O V] | 1218 Å | |
| N V | 1239 Å | |
| N V | 1240 Å | |
| N V | 1243 Å | |
| Si II | 1263 Å | |
| O I | 1304 Å | |
| Si II | 1308 Å | |
| Si IV | 1397 Å | |
| O IV] | 1402 Å | |
| S IV | 1406 Å | |
| N IV | 1485 Å | |
| N IV | 1486 Å | |
| Si II | 1531 Å | |
| C IV | 1549 Å | |
| He II | 1640 Å | |
| O III | 1661 Å | |
| O III] | 1665 Å | |
| O III | 1666 Å | |
| Al II | 1671 Å | |
| N IV | 1719 Å | |
| N III] | 1750 Å | |
| Al III | 1860 Å | |
| Si III] | 1888 Å | |
| Si III | 1892 Å | |
| C III] | 1907 Å | |
| TOTL | 1909 Å | C III] 1908.73 + [C III] |
| C III | 2297 Å | |
| [O III] | 2321 Å | |
| C II] | 2326 Å | |
| Si II] | 2335 Å | |
| [O II] | 2471 Å | |
| Al II] | 2665 Å | |
| Mg II | 2798 Å | |
| Mg II | 2803 Å | |

| Emission line | Wavelength | Notes |
|---|---|---|
| [Ne III] | 3343 Å | |
| [Ne V] | 3426 Å | |
| Ba C 0 | | Balmer Cont. |
| Balmer | 3646 Å | Ba Jump |
| [O II] | 3726 Å | |
| [O II] | 3727 Å | |
| [O II] | 3729 Å | |
| [Ne III] | 3869 Å | |
| H I | 3889 Å | |
| Ca II | 3933 Å | |
| He I | 4026 Å | |
| [S II] | 4070 Å | |
| [S II] | 4070 Å | |
| [S II] | 4078 Å | |
| H I | 4102 Å | |
| Ni 12 | 4231 Å | |
| H I | 4340 Å | |
| [O III] | 4363 Å | |
| He II | 4686 Å | |
| Ca B | 4686 Å | Case B He II |
| [Ar IV] | 4711 Å | |
| [Ne IV] | 4720 Å | |
| [Ar IV] | 4740 Å | |
| Incident | 4860 Å | |
| Hβ | 4861 Å | |
| [O III] | 4959 Å | |
| [O III] | 5007 Å | |
| Co 11 | 5168 Å | |
| [N I] | 5200 Å | |
| Fe 14 | 5303 Å | |
| Ar 10 | 5534 Å | |
| [O I] | 5577 Å | |
| [N II] | 5755 Å | |
| He I | 5876 Å | |
| [O I] | 6300 Å | |
| [S III] | 6312 Å | |
| [O I] | 6363 Å | |
| Hα | 6563 Å | |
| [N II] | 6584 Å | |
| [S II] | 6716 Å | |
| [S II] | 6720 Å | |
| [S II] | 6731 Å | |
| Ar V | 7005 Å | |
| [Ar III] | 7135 Å | |
| [O II] | 7325 Å | |
| [Ar IV] | 7331 Å | |
| [Ar III] | 7751 Å | |
| Mn 9 | 7968 Å | |
| O I | 8446 Å | |
| Ca II | 8498 Å | |
| Ca II | 8542 Å | |
| Ca II | 8662 Å | |
| Ca II | 8579 Å | |
| [S III] | 9069 Å | |
| Pa 9 | 9229 Å | |
| [S III] | 9532 Å | |
| Pa ε | 9546 Å | |
| S 8 | 9914 Å | |
| H I | 1.005 μm | |
| He I | 1.083 μm | |
| H I | 1.094 μm | |
| H I | 1.282 μm | |
| H I | 1.875 μm | |
| H I | 2.625 μm | |
| H I | 4.051 μm | |
| Na III | 7.320 μm | |
| Ne VI | 7.652 μm | |
| Ne II | 12.81 μm | |
| [Ne V] | 14.3 μm | |
| Ne III | 15.55 μm | |
| Ne V | 24.31 μm | |
| O IV | 25.88 μm | |
| Ne III | 36.01 μm | |
| O III | 51.80 μm | |
| [N III] | 57.2 μm | |
| [O I] | 63 μm | |
| [O III] | 88 μm | |
| N II | 121.7 μm | |
| [O I] | 145.5 μm | |
| C II | 157.6 μm | |
| N II | 205.4 μm | |
| Cr 8 | 1.011 m | |
| S 9 | 1.252 m | |
| V 7 | 1.304 m | |
| S 11 | 1.393 m | |
| Si 10 | 1.430 m | |
| Ti 6 | 1.715 m | |
| H I | 1.945 m | |
| S 11 | 1.920 m | |
| Si 6 | 1.963 m | |
| H I | 2.166 m | |
| Sc V | 2.310 m | |
| Ca 8 | 2.321 m | |
| Si 7 | 2.481 m | |
| Si 9 | 2.584 m | |
| Ar 11 | 2.595 m | |
| Al 5 | 2.905 m | |
| Mg 8 | 3.030 m | |
| Ca IV | 3.210 m | |
| Al 6 | 3.660 m | |
| Al 8 | 3.690 m | |
| S 9 | 3.754 m | |
| Si 9 | 3.929 m | |
| Ca V | 4.157 m | |
| Mg 4 | 4.485 m | |
| Ar 6 | 4.530 m | |
| Mg 7 | 5.503 m | |
| Mg 5 | 5.610 m | |
| Al 8 | 5.848 m | |
| Si 7 | 6.492 m | |
| Ar II | 6.980 m | |
| Ar V | 8.000 m | |
| Na 6 | 8.611 m | |
| Ar III | 9.000 m | |
| Mg 7 | 9.033 m | |
| Na 4 | 9.039 m | |
| Al 6 | 9.116 m | |
| S IV | 10.51 m | |
| Ca V | 11.48 m | |
| Ar V | 13.10 m | |
| Mg 5 | 13.52 m | |
| Na 6 | 14.40 m | |
| S III | 18.67 m | |
| Na 4 | 21.29 m | |
| Ar III | 21.83 m | |
| S III | 33.47 m | |
| Si II | 34.81 m | |